\documentclass[universe,article,accept,moreauthors,pdftex]{Definitions/mdpi} 

\firstpage{1} 
\pubvolume{7}
\issuenum{4}
\articlenumber{101}
\pubyear{2021}
\copyrightyear{2021}
\externaleditor{{Academic Editor: James A. Isenberg, Gerald B. Cleaver, Lijing Shao, Gonzalo J. Olmo and Giacomo Tommei}} 
\datereceived{30 March 2021} 
\dateaccepted{12 April 2021} 
\datepublished{15 April 2021} 
\hreflink{https://doi.org/\linebreak 10.3390/universe7040101} 

\usepackage{latexsym}
\usepackage{graphics}
\usepackage{amssymb}
\usepackage{bm}
\newcommand{\be}[1]{\begin{equation}\label{#1}}
\newcommand{\ee}{\end{equation}}
\newcommand{\ba}[1]{\begin{eqnarray}\label{#1}}
\newcommand{\ea}{\end{eqnarray}}
\newcommand{\rf}[1]{(\ref{#1})}
\newcommand{\nn}{\nonumber}

\newcommand{\const}{\mbox{\rm const}\,}

\Title{Gravitational Interaction in the Chimney Lattice Universe $^{\dagger}$}

\TitleCitation{Gravitational Interaction in the Chimney Lattice Universe}

\Author{Maxim Eingorn $^{1,}$*, Andrew McLaughlin II $^{1}$, Ezgi Canay $^{2}$\orcidA{}, Maksym Brilenkov $^{3}$ and Alexander Zhuk $^{4}$\orcidB{}}

\AuthorNames{Maxim Eingorn, Andrew McLaughlin II, Ezgi Canay, Maksym Brilenkov and Alexander Zhuk}

\AuthorCitation{Eingorn, M.; \linebreak McLaughlin, A., II; Canay, E.; Brilenkov M.; Zhuk A.}

\address{%
$^{1}$ \quad Department of Mathematics and Physics, North Carolina Central University, 1801 Fayetteville St., \mbox{Durham, NC 27707, USA}; amclau12@eagles.nccu.edu\\
$^{2}$ \quad  Department of Physics, Istanbul Technical University, Maslak, 34469 Istanbul, Turkey; ezgicanay@itu.edu.tr\\
$^{3}$ \quad Institute of Theoretical Astrophysics, University of Oslo, P.O. Box 1029 Blindern, N-0315 Oslo, Norway; maksym.brilenkov@astro.uio.no\\
$^{4}$ \quad  Astronomical Observatory, Odessa I.I. Mechnikov National University, Dvoryanskaya St. 2, \mbox{65082 Odessa, Ukraine}; ai.zhuk2@gmail.com}

\corres{Correspondence: maxim.eingorn@gmail.com or meingorn@nccu.edu}
\firstnote{This paper is an extended version from the proceeding paper:
Eingorn, M.; McLaughlin, A., II; Canay, E.; Brilenkov, M.; Zhuk, A. Gravitation in the space with chimney topology. In Proceedings of the 1st Electronic Conference on
Universe, Online, 22--28  February 2021.} 

\abstract{We investigate the influence of the chimney topology $T\times T\times  R$ of the Universe on the gravitational potential and force that are generated by point-like massive bodies. We obtain three distinct expressions for the solutions. One follows from Fourier expansion of delta functions into series using periodicity in two toroidal dimensions. The second one is the summation of solutions of the Helmholtz equation, for a source mass and its infinitely many images, which are in the form of Yukawa potentials.  The third alternative solution for the potential is formulated via the Ewald sums method applied to Yukawa-type potentials. We show that, for the present Universe, the formulas involving plain summation of Yukawa potentials 
are preferable for computational purposes, as they require a smaller number of terms in the series to reach adequate precision.}



\begin{document}

\section{Introduction}

The shape of the space, whether it is positively curved, negatively curved, or flat, and whether there is a limit to the size of the Universe are all among essential topics of contemporary debate in theoretical physics and cosmology. Spatial topology of the Universe, its function at the very early stages of evolution (in the quantum gravity regime), and in the later process of large scale structure formation are quite interesting questions yet to be answered.      
General Relativity does not favor any particular topology; hence, on theoretical grounds, the space might be simply connected, in agreement with concordance cosmology, or equally as well, multiply connected. {It is worth noting that some ``exotic'' non-simply connected spacetimes are timelike geodesically incomplete, since they have singularities \cite{Gannon}. Hence, such topologies are not viable.}

If the Universe is multiply connected, it may have a finite volume and yet be negatively curved or flat \cite{36}.  The current available data cannot reveal the finiteness of its volume if the Universe covers a much wider region than the observable sector. However, a rather smaller volume points at the possibility of finding observational indications of its topological features \cite{37}. For instance, a photon can travel plenty of times across the volume of multiply connected space and, thus, generate multiple images of the emitting source as a \mbox{signature \cite{43,44}}. Spaces with toroidal topology in one to three dimensions may be presented as common examples of multiply connected spaces. To this class belongs the three-torus $T\times T\times T$, chimney $T\times T\times R$, and slab $T\times R\times R$ topologies.  

There are various comprehensive studies in the literature on potential indicators of the shape of the space \cite{38,39,40,41,42}, and the majority of research is focused on their relation to the Cosmic Microwave Background (CMB) data. Indeed, there exists a very appealing  conjectural relation which suggests that CMB anomalies in large angular scale observations, e.g., the suppression of the quadrupole moment and the quadrupole and octopole alignment, are imprints of spatial topology \cite{47,48}. {The weak wide-angle temperature correlations in the CMB can be also explained, e.g., with the help of dodecahedral topology of the Universe \cite{Luminet}.} In the present work we study the chimney topology, which admits a single infinite axis subject to interpretations such as the preferred direction of the quadrupole and octopole alignment and the commonly named “axis of evil” \cite{49} (see \cite{51} for additional observable signatures of a preferred axis).

In connection with the investigation of possible topological imprints in CMB observations, Planck 2013 data \cite{36} place the constraint $R_i>0.71 \chi_{\mathrm{rec}}$ on the radius of the largest sphere that may be inscribed in the topological domain for a flat Universe with the equal-sided chimney topology. The parameter $\chi_{\mathrm{rec}}$ specifies the distance from the recombination surface and it is of the same order with the particle horizon, that is, $\chi_{\mathrm{rec}}\sim 14\, \rm Gpc$. For the toroidal topologies, the former restrictions on the size of the Universe from the seven- and nine-year WMAP temperature map analyses are presented in \cite{49,52}. The smallest possible size of the fundamental  topological domain for flat space, according to the seven-year WMAP results, is $d=2R_{\mathcal{LSS}}\cos\left(\alpha_{\mathrm{min}}\right)\simeq27.9\, \rm Gpc$ \cite{48}, where $R_{\mathcal{LSS}}$ stands for the distance from the last scattering surface.    

In this paper, we study the chimney topology $T\times T\times R$ in terms of the gravitational characteristics of the Universe, manifested in the shape of the gravitational potential and force. In the cosmological setting, the inhomogeneous gravitational field is sourced by fluctuations in the matter density \cite{Peebles} and, as expected, in the Newtonian limit, the potential satisfies the Poisson equation. The form of the gravitational potential in the case of toroidal topologies was previously studied in \cite{topology1}. Particularly, the authors have shown that there exists no physically justified nontrivial solution of the Poisson equation for the $T\times T\times R$ model. On the other hand, by employing the perturbed Einstein equations from the very beginning, one automatically includes the essential relativistic effects in the formulation and, for the gravitational potential, obtains a Helmholtz-type equation instead of the Poisson one \cite{Eingorn1,Claus1,Claus2}. Quite remarkably, as we show in the present work for the chimney topology, it then becomes possible to obtain exact solutions of this equation that are nontrivial and physically meaningful. Herein, we derive distinct expressions for the gravitational potential and force through alternative methods and point out the particular solutions appearing in the form of summed Yukawa potentials as ready-to-use notable sources for numerical computations. It is worth mentioning that, in the above outlined approach, we make no presumptions regarding the spatial distribution of gravitating~bodies.

The outline of the paper is as follows. In Section~\ref{sec:2}, following \cite{proceedings}, we introduce the main equations and derive alternative expressions for the gravitational potential using distinct methods, now including the Ewald technique. Subsequently, in Section~\ref{sec:3}, extending the results of \cite{proceedings}, we compare these expressions in view of their usefulness for numerical computations. In Section~\ref{sec:4}, we obtain the gravitational force expressions for each form of the potential. We briefly review the results of our work in the concluding Section~\ref{sec:5}.
 
\section{Methods}\label{sec:2}
\subsection*{The Model and Basic Equations}
It is well known that the gravitational potential $\Phi$, created by fluctuations in the matter density, is defined by scalar perturbations of the metric coefficients 
\cite{Landau} and that in the framework of General Relativity, it satisfies the linearized Einstein equation (see, \mbox{e.g., \cite{Mukhanov2,Rubakov}}). Ignoring peculiar velocities, in the case 
of the $\Lambda$CDM cosmological model, this equation reads \cite{Eingorn1,Claus1,Claus2}
\be{2.1}
\Delta\Phi_0 - \frac{3\kappa\overline{\rho} c^2}{2a}{\Phi}_0=\frac{\kappa c^2}{2a}\left(\rho -\bar{\rho}\right)\, ,
\ee
where $\kappa\equiv 8\pi G_N/c^4$ (with $G_N$ and $c$ being the Newtonian gravitational constant and the speed of light, respectively), $a$ is the scale factor, while $\Delta$ represents the Laplace operator in comoving coordinates. Here, $\rho$ and $\bar{\rho}=\const$ are the comoving mass density and its averaged value, respectively. As we operate within the $\Lambda$CDM model, matter is pressureless, and we  consider it in the form of discrete point-like gravitating bodies with masses $m_n$ to represent, e.g., galaxies.
Therefore, the comoving mass density 
\be{2.2}
\rho = \sum_n m_n\delta(\mathbf{r}-\mathbf{r}_n)\, .
\ee

The $0$ subscript of $\Phi$ in  Equation~\rf{2.1} refers to the  fact that peculiar velocities have been disregarded (see also \cite{cosmlaw}).

The shifted gravitational potential 
\be{2.3}
\widehat{\Phi}_0\equiv \Phi_0 -\frac13
\ee
fulfils the equation
\be{2.4}
\Delta\widehat{\Phi}_0 - \frac{a^2}{\lambda^2} \widehat{\Phi}_0=\frac{\kappa c^2}{2a}\rho\, , 
\ee
where the screening length \cite{Eingorn1}
\be{2.5}
\lambda \equiv \left(\frac{3\kappa\overline{\rho} c^2}{2a^3}\right)^{-1/2}\, .
\ee

{The presence of the term $\propto \Phi_0$ in  Equation~\rf{2.1} (consequently, $\propto \widehat{\Phi}_0$ in  Equation~\rf{2.4}) results in the Yukawa-type cutoff of the potential with the characteristic length $\lambda$. The term $\propto \Phi$ enters as a summand into energy-momentum fluctuations generating metric perturbations~\cite{Eingorn1}.}

In what follows, the overhat indicates that the gravitational potential is shifted. A significant bonus of working with the shifted potential is that it is now possible to employ the superposition principle in solving  Equation~\rf{2.4}: once we find a solution for a single particle that is located at the center of Cartesian coordinates, we may immediately generalize it for a collection of particles at random positions.

We consider the space with chimney topology  $T_1\times T_2 \times R$, where the tori $T_1$ and $T_2$ have periods $l_1$ and $l_2$  along, e.g., the $x$- and $y$-axes, respectively. Hence, each gravitating body has its images positioned away from the original point in multiples of periods $l_1$ and $l_2$ along the corresponding axes. Now, let us place a particle with mass $m$ at the center of Cartesian coordinates. For the above indicated topology, the delta functions $\delta(x)$ and $\delta(y)$ may be presented as
\be{2.6} \delta(x)=\frac{1}{l_1}\sum_{k_1=-\infty}^{+\infty}\cos\left(\frac{2\pi k_1}{l_1}x\right)\, ,\quad 
\delta(y)=\frac{1}{l_2}\sum_{k_2=-\infty}^{+\infty}\cos\left(\frac{2\pi k_2}{l_2}y\right)\, , 
\ee
which implicitly include the contribution from the  images of the particle.
Consequently,  Equation~\rf{2.4} for this particle reads
\ba{2.7}
&&\Delta\widehat{\Phi}_0-\frac{a^2}{\lambda^2}\widehat{\Phi}_0=\frac{\kappa
	c^2}{2a}\frac{m}{l_1l_2}\sum_{k_1=-\infty}^{+\infty}\sum_{k_2=-\infty}^{+\infty} \cos\left(\frac{2\pi k_1}{l_1}x\right)\cos\left(\frac{2\pi
	k_2}{l_2}y\right)\delta\left(z\right)\, ,
\ea
so, we are motivated to consider the solution 
\be{2.8} \widehat{\Phi}_0=\sum_{k_1=-\infty}^{+\infty}\sum_{k_2=-\infty}^{+\infty} C_{k_1k_2}(z)\cos\left(\frac{2\pi
	k_1}{l_1}x\right)\cos\left(\frac{2\pi k_2}{l_2}y\right)\, ,\ee
where the coefficients $C_{k_1k_2}(z)$ satisfy the equation
\ba{2.9}
&&\sum_{k_1=-\infty}^{+\infty}\sum_{k_2=-\infty}^{+\infty}\left[C^{''}_{k_1k_2}(z)-4\pi^2\left(\frac{k_1^2}{l_1^2}+
\frac{k_2^2}{l_2^2}\right)C_{k_1k_2}(z)-\frac{a^2}{\lambda^2}C_{k_1k_2}(z)-\frac{\kappa
	c^2}{2a}\frac{m}{l_1l_2}\delta\left(z\right)\right]\nn\\
&\times&\cos\left(\frac{2\pi k_1}{l_1}x\right)\cos\left(\frac{2\pi k_2}{l_2}y\right) =0\, .
\ea

Using the condition $d^2|z|/dz^2=2\delta(z)$, we can easily obtain the explicit expressions for the coefficients $C_{k_1k_2}(z)$, so that the shifted gravitational potential for the selected particle and all its images eventually reads 
\ba{2.10} 
\widehat{\Phi}_0&=&-\frac{\kappa c^2}{4a}\frac{m}{l_1l_2}\sum_{k_1=-\infty}^{+\infty}\sum_{k_2=-\infty}^{+\infty} \left[4\pi^2\left(\frac{k_1^2}{l_1^2}+\frac{k_2^2}{l_2^2}\right)+
\frac{a^2}{\lambda^2}\right]^{-1/2}\nn\\
&\times& \exp\left(-\sqrt{4\pi^2\left(\frac{k_1^2}{l_1^2}+\frac{k_2^2}{l_2^2}\right)+\frac{a^2}{\lambda^2}}|z|\right)\cos\left(\frac{2\pi k_1}{l_1}x\right)\cos\left(\frac{2\pi k_2}{l_2}y\right)\, . 
\ea

The above expression has the correct behavior in the Newtonian limit in the neighborhood of the considered particle, where it is no longer possible to distinguish between different (infinite and periodic) axes. For such regions, the summations in \rf{2.10} may be replaced by the integrals:
\ba{2.11} 
\widehat{\Phi}_0&\rightarrow&-\frac{\kappa c^2m}{4a}\int_{-\infty}^{+\infty}dk_x\int_{-\infty}^{+\infty}dk_y \left[4\pi^2\left(k_x^2+k_y^2\right)+
\frac{a^2}{\lambda^2}\right]^{-1/2}\nn\\
&\times&\exp\left(-\sqrt{4\pi^2\left(k_x^2+k_y^2\right)+\frac{a^2}{\lambda^2}}|z|\right)\cos\left[2\pi\left(
k_xx+k_y y\right)\right]\, 
\ea
for $k_x\equiv k_1/l_1$ and $k_y\equiv k_2/l_2$. Introducing the vectors ${\bf{k}}=(k_x,k_y)$, ${\bf{r}}=(x,y)$ with the absolute values $k=\sqrt{k_x^2+k_y^2}$ and $r=\sqrt{x^2+y^2}$, and assuming an angle $\varphi$ between them, we obtain
\ba{2.12}
\widehat{\Phi}_0&\rightarrow&-\frac{\kappa c^2m}{4a}\int_{0}^{+\infty}kdk \left[4\pi^2k^2+\frac{a^2}{\lambda^2}\right]^{-1/2}\exp\left(-\sqrt{4\pi^2k^2+\frac{a^2}{\lambda^2}}|z|\right)\,\nn\\
&&\times\int_{0}^{2\pi}d\varphi
\cos\left(2\pi kr\cos\varphi\right)\nn\\
&=&-\frac{\pi\kappa c^2m}{2a}\int_{0}^{+\infty}kdk \left[4\pi^2k^2+ \frac{a^2}{\lambda^2}\right]^{-1/2}\exp\left(-\sqrt{4\pi^2k^2+\frac{a^2}{\lambda^2}}|z|\right)J_0(2\pi kr)\nn\\
&=& -\frac{G_N m}{c^2} \frac{1}{\sqrt{Z^2+R^2}}\exp\left(-\frac{1}{\lambda}\sqrt{Z^2+R^2}\right)
\rightarrow -\frac{G_N m}{c^2} \frac{1}{\sqrt{Z^2+R^2}}
\, ,
\ea
where $Z=az$ and $R=ar$ represent the physical distances, and the last integration is performed by using the formula 2.12.10(10) of \cite{Table}.

Evidently, for a system of randomly positioned gravitating bodies, we have
\be{2.13} 
\Phi_0=\frac{1}{3}+\widehat{\Phi}_0\left\{m\rightarrow\sum_nm_n;\ x,y,z\rightarrow x-x_n,y-y_n,z-z_n\right\}\, .\ee

For linear fluctuations, the averaged value of this expression is equal to zero, as it should be  (also see \cite{EBV}). Indeed,
\ba{2.14} &&\int\limits_{0}^{l_1}dx\int\limits_{0}^{l_2}dy\int\limits_{-\infty}^{+\infty}dz\widehat{\Phi}_0=-\frac{\kappa c^2}{4a}m\frac{\lambda}{a}\int\limits_{-\infty}^{+\infty}\exp{\left(-\frac{a}{\lambda}|z|\right)}dz\nn\\
&=&\frac{\kappa c^2}{2a}m\frac{\lambda^2}{a^2}\left.\exp{\left(-\frac{a}{\lambda}z\right)}\right|_{0}^{+\infty}=-\frac{\kappa c^2}{2a}m\frac{\lambda^2}{a^2}=-\frac{1}{3}\frac{m}{\overline\rho}\, ,
\ea
and, hence, the spatial average of the total gravitational potential is 
\be{2.15} \overline{\Phi_0}=\frac{1}{3}-\frac{1}{3}\frac{m}{\overline\rho}\cdot \frac{N}{l_1l_2L_z}=\frac{1}{3}-\frac{1}{3}=0,\quad \frac{m N}{l_1l_2
	L_z}=\overline\rho\, .
\ee

For the sake of simplicity, here we have considered the particular configuration in which all $N$ bodies in the volume $V = l_1l_2L_z$ are assigned identical masses $m$.

 Equation~\rf{2.4} is of Helmholtz type and we can likewise solve it by considering the contribution of periodic images. In this case, the resulting expression consists of summed Yukawa potentials attributed to each one of them:
\ba{2.16} 
\widehat{\Phi}_{0}&=&-\frac{\kappa c^2m}{8\pi a}\sum_{k_1=-\infty}^{+\infty}\sum_{k_2=-\infty}^{+\infty} \frac{1}{\sqrt{(x-k_1l_1)^2+(y-k_2l_2)^2+z^2}}\nn\\
&\times&\exp\left(-\frac{a\sqrt{(x-k_1l_1)^2+(y-k_2l_2)^2+z^2}}{\lambda}\right)\, .
\ea
 
As we have noted previously, the peculiar motion of gravitating bodies is disregarded in  Equation~\rf{2.1} and, consequently, in~\rf{2.4}. Nevertheless, the significance of such a contribution has recently been pointed out in \cite{MaxEzgi}, where the authors have also shown that peculiar velocities may be effectively restored by employing the effective screening length $\lambda_{\mathrm{eff}}$ (given by the Formula (41) of \cite{MaxEzgi}) instead of the screening length $\lambda$ in  \mbox{Equations~\rf{2.1} and~\rf{2.4}}. Specifically, in the matter-dominated epoch, the two quantities $\lambda_{\mathrm{eff}}$ and $\lambda$ are related to one another as $\lambda_{\mathrm{eff}}=\sqrt{3/5}\lambda$. Returning to our formulation, the effect of peculiar motion is included by replacing $\lambda$ with $\lambda_{\mathrm{eff}}$ in the Formulas \rf{2.10} and \rf{2.16}, which yields
\ba{2.17} 
\tilde{\Phi}_{\cos}&\equiv& \left(-\frac{\kappa c^2}{8\pi a}\frac{m}{l}\right)^{-1}\widehat{\Phi}_{\cos}=\sum_{k_1=-\infty}^{+\infty}\sum_{k_2=-\infty}^{+\infty} \left(k_1^2+k_2^2+
\frac{1}{4\pi^2\tilde\lambda_{\mathrm{eff}}^2}\right)^{-1/2}\nn\\
&\times& \exp\left(-\sqrt{4\pi^2\left(k_1^2+k_2^2\right)+ \frac{1}{\tilde\lambda_{\mathrm{eff}}^2}}|\tilde z|\right)\cos\left(2\pi k_1\tilde x\right)\cos\left(2\pi k_2\tilde y\right)\,  
\ea
and
\ba{2.18} 
\tilde{\Phi}_{\exp}&\equiv&\left(-\frac{\kappa c^2}{8\pi a}\frac{m}{l}\right)^{-1}\widehat{\Phi}_{\exp}=\sum_{k_1=-\infty}^{+\infty}\sum_{k_2=-\infty}^{+\infty} \frac{1}{\sqrt{\left(\tilde x-k_1\right)^2+\left(\tilde y-k_2\right)^2+\tilde z^2}}\nn\\
&\times&\exp\left(-\frac{\sqrt{\left(\tilde x-k_1\right)^2+\left(\tilde y-k_2\right)^2+\tilde z^2}}{\tilde\lambda_{\mathrm{eff}}}\right)\, . \ea
\textls[-25]{For simpler demonstration, we have assumed $l_1=l_2=l$ and introduced the rescaled quantities} 
\be{2.19} 
x=\tilde xl,\quad y=\tilde yl,\quad z=\tilde zl,\quad \lambda_{\mathrm{eff}}=\tilde\lambda_{\mathrm{eff}}al\, . 
\ee
Two alternative solutions are labeled with the subscripts ``cos'' and ``exp'' in \mbox{ Equations~\rf{2.17}} and \rf{2.18}. Now that the peculiar velocities are included in the calculations, the $0$ subscript is omitted in the new formulas.

There is also a third way to express the gravitational potential for the given topology. 
Indeed, Yukawa-type interactions that are subject to periodic boundary conditions can be formulated via Ewald sums, so that the expression for the potential
consists of two rapidly converging series, one in each of the real and 
Fourier spaces. The technique is commonly employed while modeling 
particle interactions in plasma and
colloids, and, in such a context, the corresponding potential for 
quasi two-dimensional systems, i.e., three-dimensional systems with 
two-dimensional periodicity, has previously been derived in 
\cite{Mazars1,Mazars2}. Being implemented in the cosmological 
setting considered in our paper, the discussed expression for the ``Yukawa--Ewald'' potential reads
\ba{2.20} 
&&\tilde{\Phi}_{\mathrm{mix}}\equiv\left(-\frac{\kappa c^2}{8\pi a}\frac{m}{l}\right)^{-1}\widehat{\Phi}_{\mathrm{mix}}\,\nn\\
&=&\sum_{k_1=-\infty}^{+\infty}\sum_{k_2=-\infty}^{+\infty}\left[\frac{D\left(\sqrt{\left(\tilde x-k_1\right)^2+\left(\tilde y-k_2\right)^2+\tilde z^2};\alpha;\tilde\lambda_{\mathrm{eff}}\right)}{2\sqrt{\left(\tilde x-k_1\right)^2+\left(\tilde y-k_2\right)^2+\tilde z^2}}\right. \,\nn\\
&+&\left.\pi \cos\left[2\pi \left(k_1\tilde x+k_2\tilde y\right) \right]\frac{F\left(\sqrt{4\pi^2 (k_1^2+k_2^2)+\tilde\lambda^{-2}_{\mathrm{eff}}};\tilde z;\alpha\right)}{\sqrt{4\pi^2(k_1^2+k_2^2)+\tilde\lambda^{-2}_{\mathrm{eff}}}}\right]\, ,
\ea    
where
\ba{2.21} 
&&D\left(\sqrt{\left(\tilde x-k_1\right)^2+\left(\tilde y-k_2\right)^2+\tilde z^2};\alpha;\tilde\lambda_{\mathrm{eff}}\right)\,\nn\\
&\equiv&\exp\left(\frac{\sqrt{\left(\tilde x-k_1\right)^2+\left(\tilde y-k_2\right)^2+\tilde z^2}}{\tilde\lambda_{\mathrm{eff}}}\right)\,\nn\\
&\times&\mathrm{erfc}\left(\alpha \sqrt{\left(\tilde x-k_1\right)^2+\left(\tilde y-k_2\right)^2+\tilde z^2}+\frac{1}{2\alpha\tilde\lambda_{\mathrm{eff}}}\right)\,\nn\\
&+&\exp\left(-\frac{\sqrt{\left(\tilde x-k_1\right)^2+\left(\tilde y-k_2\right)^2+\tilde z^2}}{\tilde\lambda_{\mathrm{eff}}}\right)\,\nn\\
&\times&\mathrm{erfc}\left(\alpha \sqrt{\left(\tilde x-k_1\right)^2+\left(\tilde y-k_2\right)^2+\tilde z^2}-\frac{1}{2\alpha\tilde\lambda_{\mathrm{eff}}}\right)
\ea
and
\ba{2.22}&& F\left(\sqrt{4\pi^2(k_1^2+k_2^2)+\tilde\lambda^{-2}_{\mathrm{eff}}};\tilde z;\alpha\right)\,\nn\\
&\equiv&\exp\left(\sqrt{4\pi^2(k_1^2+k_2^2)+\tilde\lambda^{-2}_{\mathrm{eff}}}\tilde z\right)\mathrm{erfc}\left(\frac{\sqrt{4\pi^2(k_1^2+k_2^2)+\tilde\lambda^{-2}_{\mathrm{eff}}}}{2\alpha}+\alpha \tilde z\right)\,\nn\\
&+&\exp\left(-\sqrt{4\pi^2(k_1^2+k_2^2)+\tilde\lambda^{-2}_{\mathrm{eff}}}\tilde z\right)\mathrm{erfc}\left(\frac{\sqrt{4\pi^2(k_1^2+k_2^2)+\tilde\lambda^{-2}_{\mathrm{eff}}}}{2\alpha}-\alpha \tilde z\right) \, .
\ea

In these formulas, erfc is the complementary error function and the free parameter $\alpha$, as indicated in \cite{Mazars2}, is 
to be chosen in such a way that a balanced interplay of computational 
cost and satisfactory precision is achieved. For definiteness, we set $\alpha$ equal to $\tilde\lambda_{\mathrm{eff}}$.

In the forthcoming section, we will compare three expressions and present the optimum formula in view of its efficiency in use for numerical analysis.
\section{Results}
\subsection{Gravitational Potentials}\label{sec:3}
Formulas \rf{2.17}, \rf{2.18} and \rf{2.20} describe the gravitational potential due to
a point-like body, with mass $m$, placed at $(x,y,z) = (0,0,0)$, and by the accompanying images placed at $(x,y,z)=(k_1l,k_2l,0)$, where $k_{1,2}=0,\pm 1, \pm 2, \ldots$ 
All three forms of the rescaled potential are composed of infinite series. Hence, for any desired precision, one needs to determine the minimum number of terms that are needed to numerically calculate the potential.
The criterion that we use to specify 
this number $n$ is the following: the ratio $|\mathrm{exact}\ \tilde\Phi - \mathrm{approximate}\ \tilde\Phi|/|\mathrm{exact}\ \tilde\Phi|$ should be less than 0.001. This defines the order of accuracy in our analysis. Evidently, for each form of the potential the number $n$ can be different, so we denote these as $n_{\mathrm{exp}}, n_{\mathrm{cos}}$ and $n_{\mathrm{mix}}$, correspondingly. The formula requiring the smallest number of terms to define $\tilde{\Phi}$ (i.e., to calculate ``approximate $\tilde\Phi$'') up to the adopted accuracy is clearly the best alternative for numerical computation purposes.
In this connection, we are interested in comparing \rf{2.17},  \mbox {\rf{2.18}} and \rf{2.20} here. Because the formulas include double series, the accompanying numbers $n$ are to be ascribed the smallest possible number of combinations $(k_1,k_2)$ that can provide the necessary precision. We find these by listing the summands in increasing order for $\sqrt{k_1^2+k_2^2}$ and assigning to $n$ the total number of terms that are included in the list eventually. This procedure of generating a sequence of combinations $(k_1,k_2)$ and finding $n$ is performed using Mathematica \cite{Math}. 

\begin{figure}[H]
\widefigure
\resizebox{0.47\textwidth}{!}{\includegraphics{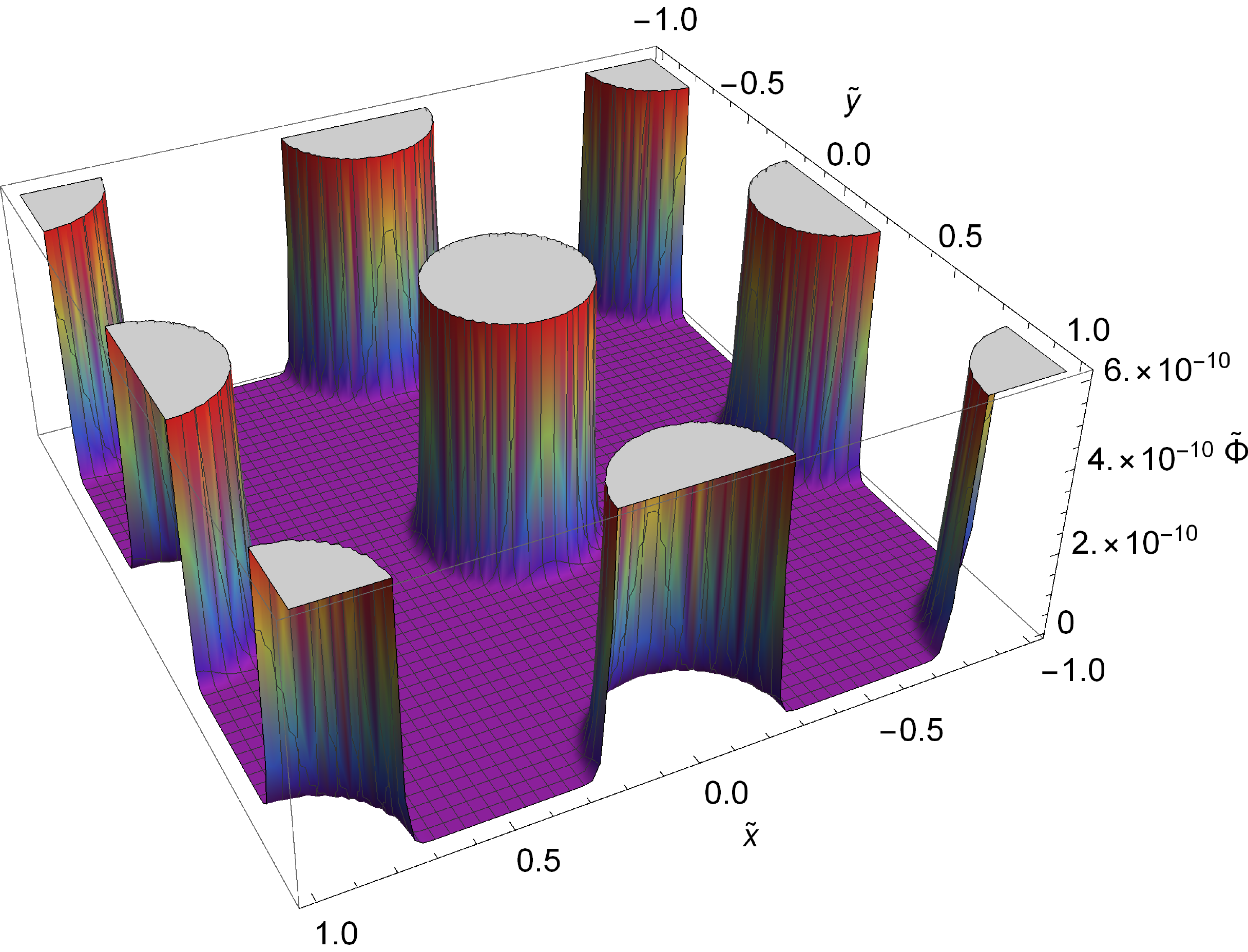}}\quad\quad
	\resizebox{0.47\textwidth}{!}{\includegraphics{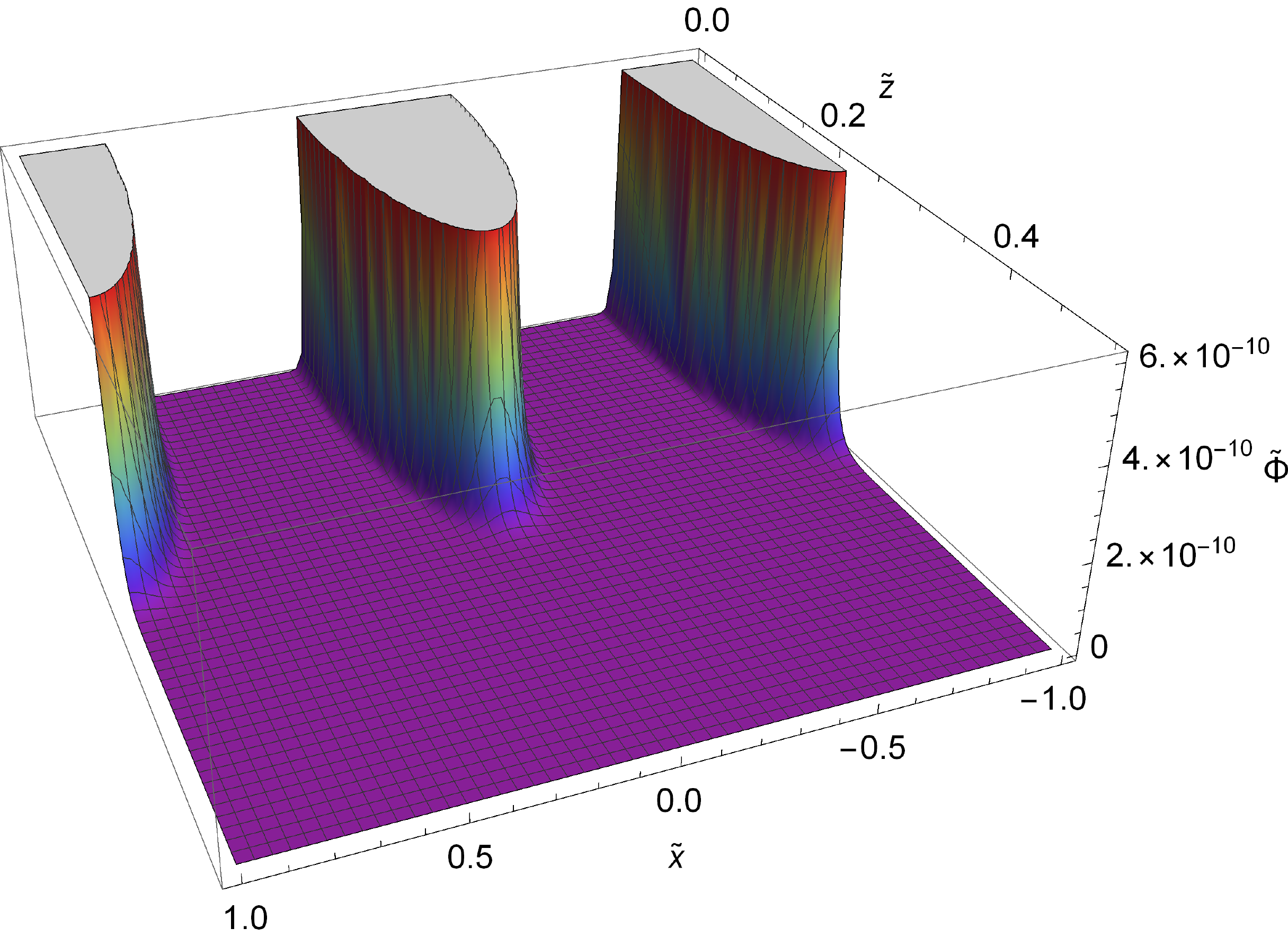}}
\caption{$\tilde\Phi=\left[-G_N m/(c^2al)\right]^{-1}\widehat\Phi$ for $\tilde\lambda_{\mathrm{eff}} = 0.01$  for the sections $z=0$ (left panel) and $y=0$ (right panel).\label{fig:1}}
\end{figure}

\begin{figure}[H]
\widefigure
\resizebox{0.47\textwidth}{!}{\includegraphics{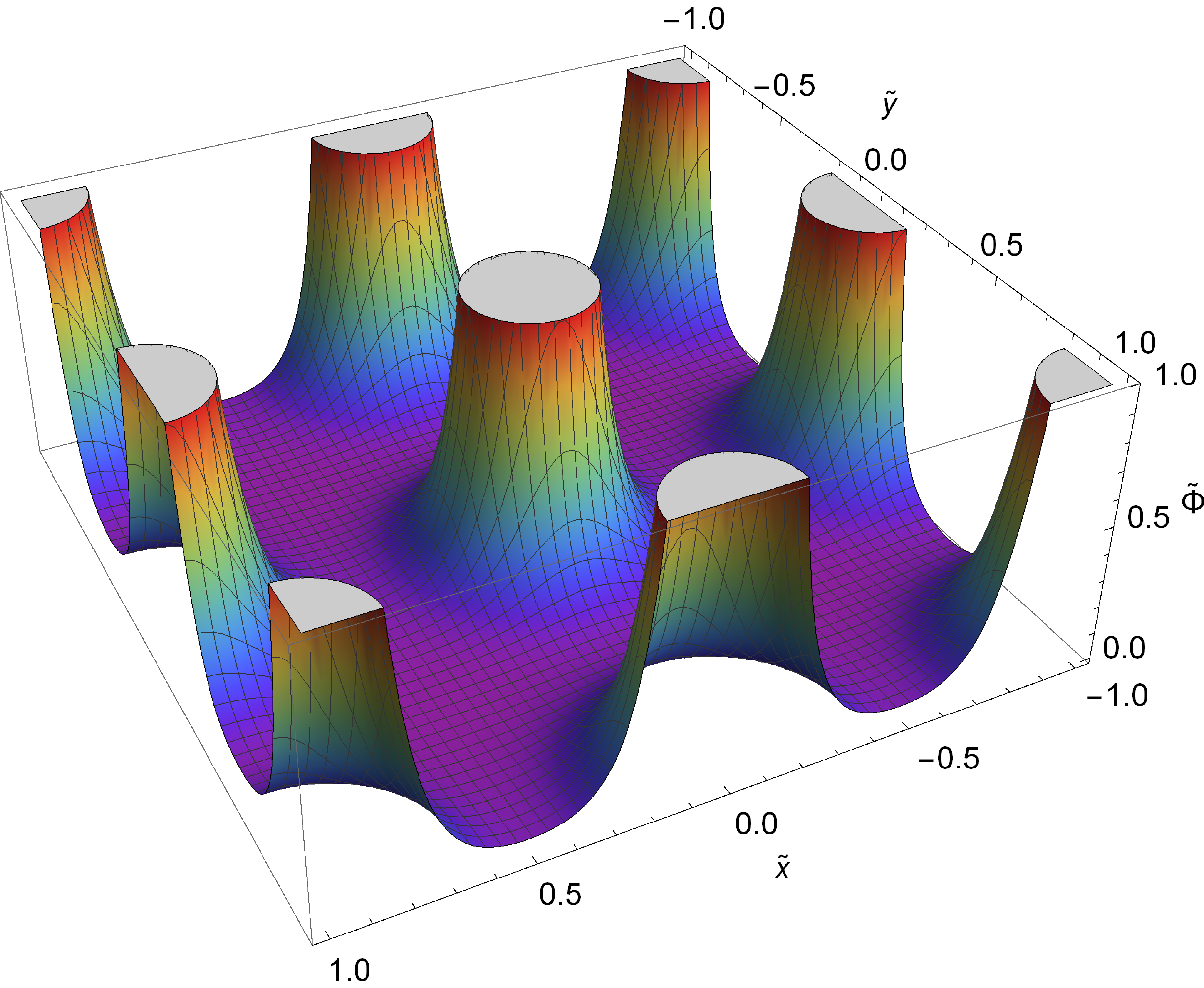}}\quad\quad
	\resizebox{0.47\textwidth}{!}{\includegraphics{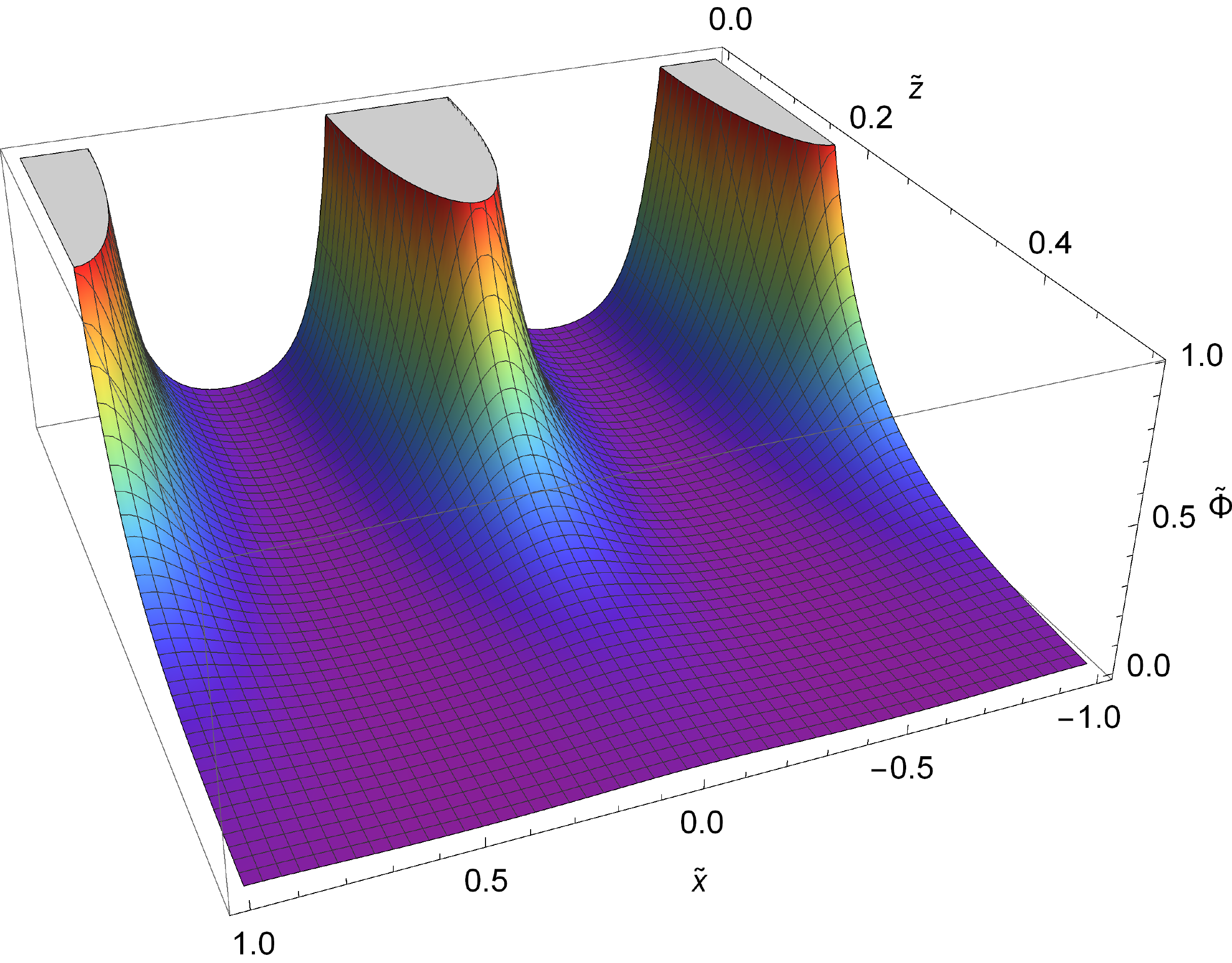}}
\caption{$\tilde\Phi=\left[-G_N m/(c^2al)\right]^{-1}\widehat\Phi$ for $\tilde\lambda_{\mathrm{eff}} = 0.1$  for the sections $z=0$ (left panel) and $y=0$ (right panel).\label{fig:2}}
\end{figure}

\begin{figure}[H]
\widefigure
\resizebox{0.47\textwidth}{!}{\includegraphics{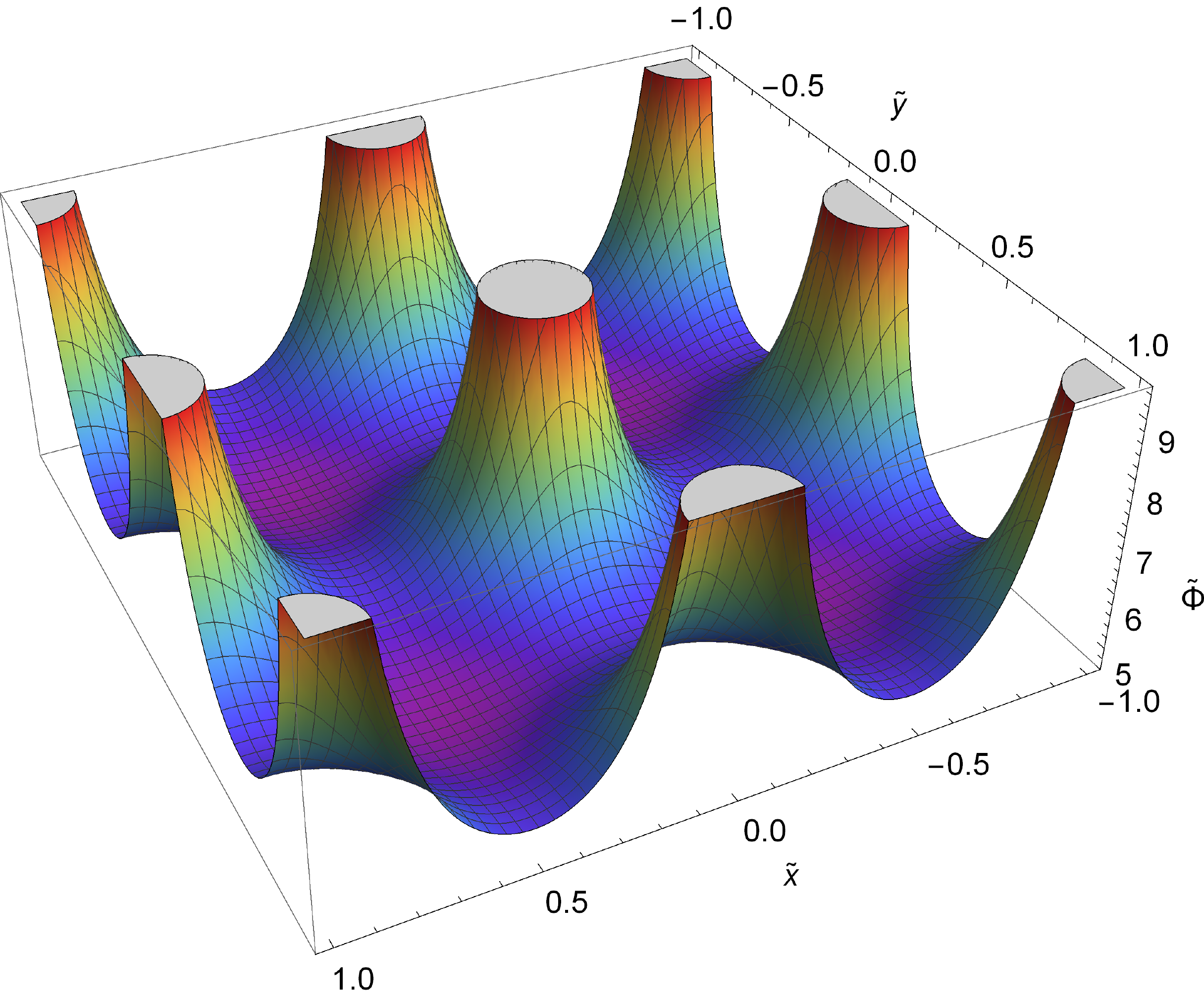}}\quad\quad
	\resizebox{0.47\textwidth}{!}{\includegraphics{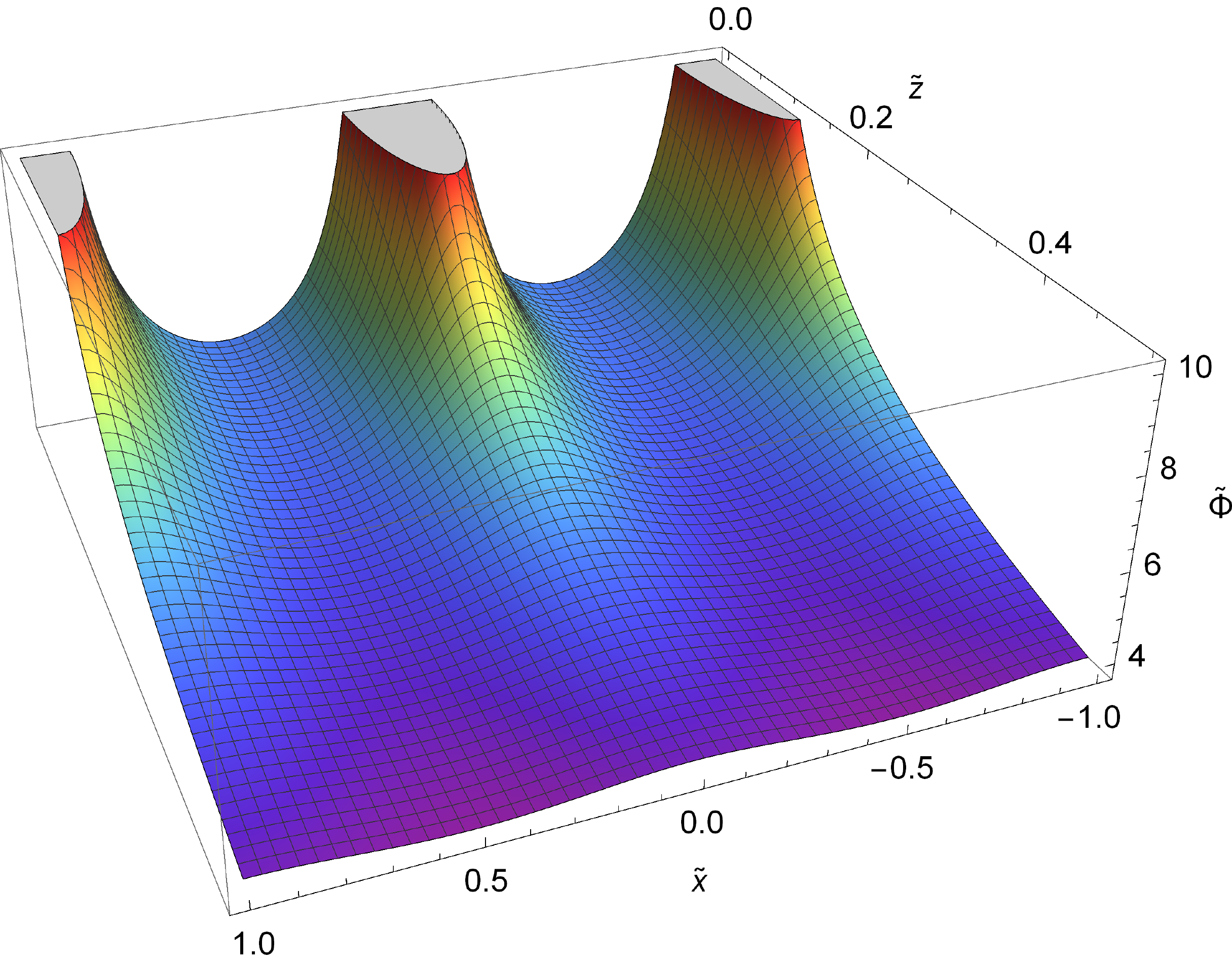}}
\caption{$\tilde\Phi=\left[-G_N m/(c^2al)\right]^{-1}\widehat\Phi$ for $\tilde\lambda_{\mathrm{eff}} = 1$  for the sections $z=0$ (left panel) and $y=0$ (right panel).\label{fig:3}}
\end{figure}

\begin{figure}[H]
\widefigure
\resizebox{0.47\textwidth}{!}{\includegraphics{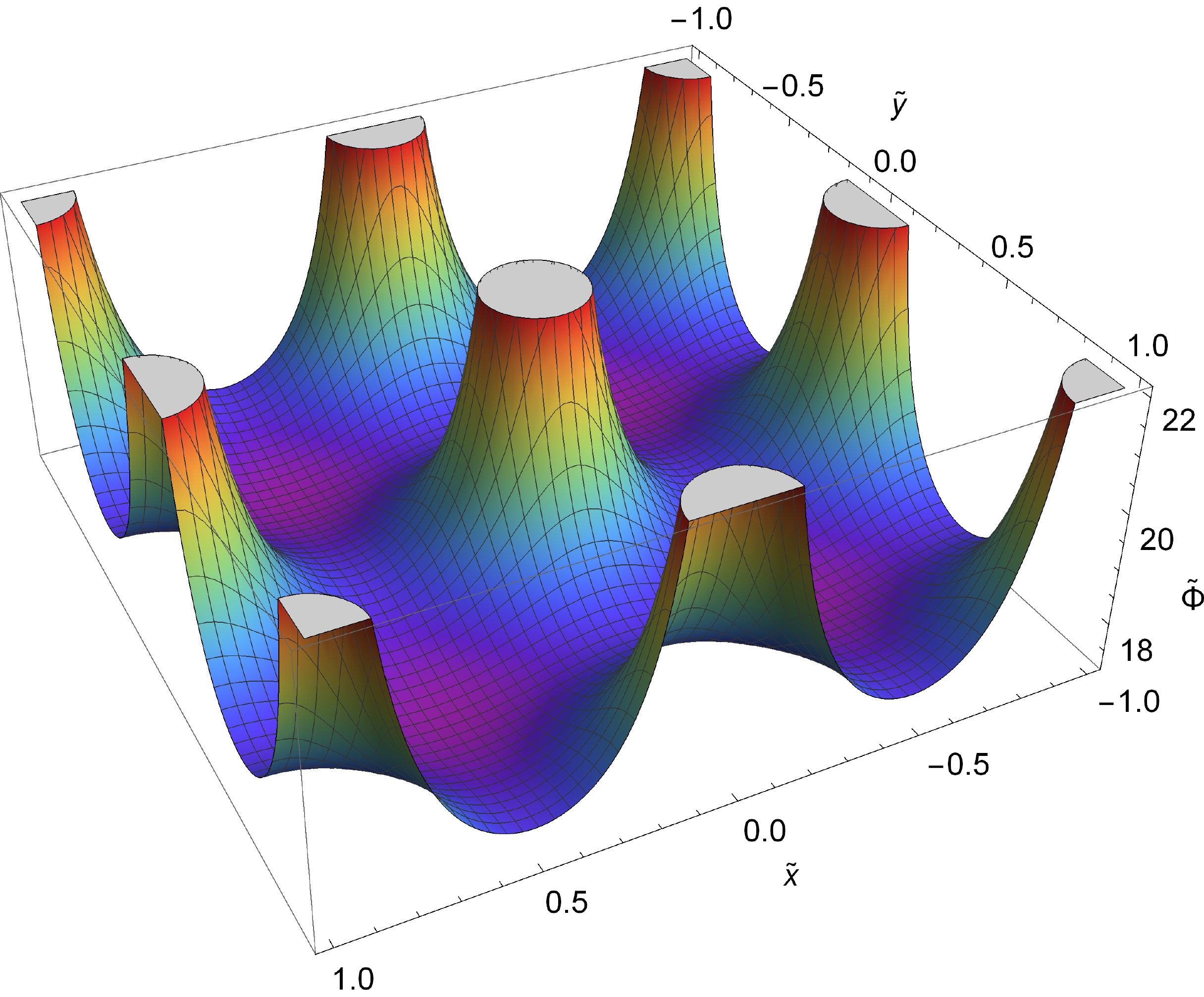}}\quad\quad
	\resizebox{0.47\textwidth}{!}{\includegraphics{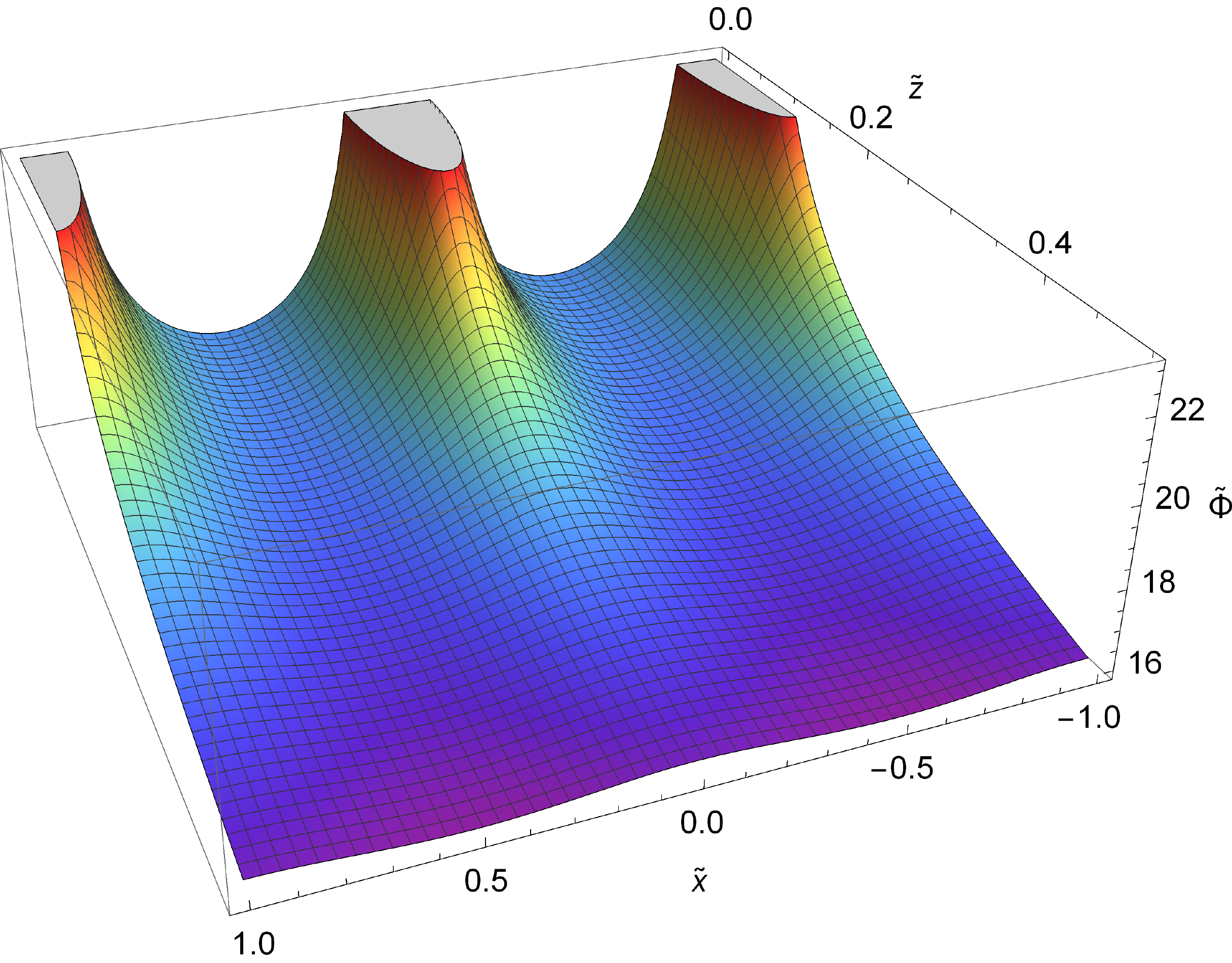}}
\caption{$\tilde\Phi=\left[-G_N m/(c^2al)\right]^{-1}\widehat\Phi$ for $\tilde\lambda_{\mathrm{eff}} = 3$  for the sections $z=0$ (left panel) and $y=0$ (right panel).\label{fig:4}}
\end{figure}

Tables~\ref{results_table_1} and~\ref{results_table_2} show the outputs for eight selected points. As to the adopted accuracy, for all $n\geqslant n_{\exp}$, the approximate value of $\tilde{\Phi}_{\exp}$ (calculated by \rf{2.18})~differs from the exact value by less than one tenth of a percent. In both tables, the exact value $\tilde\Phi$ is calculated from~\rf{2.18} for $n \gg n_{\exp}$. The quantities $n_{\cos}$ and  $n_{\mathrm{mix}}$ indicate the numbers of terms in formulas ~\rf{2.17} and \mbox{~\rf{2.20}}, respectively, which one needs to keep in order to obtain the same values of the potential at the selected points with the same precision as attained by \mbox{using~\rf{2.18}}. In the $n_{\cos}$ column, the dash reflects either incorrect outputs that are produced because of complications in the computational process, or the fact that unreasonably large number of summands is necessary. Because our results depend on the ratio of $\lambda_{\mathrm{eff}}$ to the physical size $al$ of the periods of tori, i.e., $\tilde\lambda_{\mathrm{eff}}=\lambda_{\mathrm{eff}}/(al)$, we present the results that were obtained for small and large values of $\tilde\lambda_{\mathrm{eff}}$ separately in Tables~\ref{results_table_1} and~\ref{results_table_2}, which include $\tilde\lambda_{\mathrm{eff}}=0.01,0.1$ and $\tilde\lambda_{\mathrm{eff}}=1,3$, respectively. 
\end{paracol}
\begin{specialtable}[H] 
\widetable
\caption{\label{results_table_1}Potentials $\tilde\Phi$ as well as the numbers $n_{\exp}, n_{\cos}$ and $n_{\mathrm{mix}}$ at a selection of points for $\tilde\lambda_{\mathrm{eff}}=0.01$ and $\tilde\lambda_{\mathrm{eff}}=0.1$ in the left and right tables, respectively.}
\begin{tabular*}{\hsize}{@{}@{\extracolsep{\fill}}cccccccc|cccccccc@{}}
		\noalign{\hrule height 1pt}
		& $\bm{\tilde x}$ & $\bm{\tilde y}$ & $\bm{\tilde z}$ &$\bm{\tilde{\Phi}}$ & $\bm{n_{\exp}}$ & $\bm{n_{\cos}}$ & $\bm{n_{\mathrm{mix}}}$&& $\bm{\tilde x}$ & $\bm{\tilde y}$ & $\bm{\tilde z}$ &$\bm{\tilde{\Phi}}$ & $\bm{n_{\exp}}$ & $\bm{n_{\cos}}$ & $\bm{n_{\mathrm{mix}}}$\\	
		\noalign{\hrule height 0.5pt}
	$A_1$ & 0.5 &  0 & 0.5 & $5.524\times10^{-31}$ & 2 & 1007 & 2 &$A_1$ & 0.5 & 0 & 0.5 & $2.418\times10^{-3}$ & 7 & 40& 7 \\
	$A_2$ & 0.5 & 0 & 0.1 & $2.810\times10^{-22}$ & 2 & --- & 2 &$A_2$ & 0.5 & 0 & 0.1 & $2.398\times10^{-2}$ & 6 & 808 & 6\\
	$A_3$ & 0.5 &  0 & 0 & $7.715\times10^{-22}$ & 2 &--- & 2 &$A_3$ & 0.5 & 0 & 0 & $2.700\times10^{-2}$ & 4 &--- & 4 \\
	$B_1$ & 0.1 & 0 & 0.5 & $1.405\times10^{-22}$ & 1 & 187 & 1 &$B_1$ & 0.1 & 0 & 0.5 & $1.203\times10^{-2}$ & 4 & 28 & 4 \\
	$B_2$ & 0.1 & 0 & 0.1 & $5.101\times10^{-6}$ & 1 & 2119 & 1&$B_2$ & 0.1 & 0 & 0.1 & $1.719$ & 1 & 380 & 1 \\
	$B_3$ & 0.1 & 0 & 0 & $4.540\times10^{-4}$ & 1 &--- & 1 &$B_3$ & 0.1 & 0 & 0 & $3.679$ & 1 &--- & 1\\
	$C_1$ & 0 & 0 & 0.5 & $3.857\times10^{-22}$ & 1 & 236 & 1&$C_1$ & 0 & 0 & 0.5 & $1.353\times10^{-2}$ & 4 & 37 & 4 \\
	$C_2$ & 0 & 0 & 0.1 & $4.540\times10^{-4}$ & 1 & 1479 & 1&$C_2$ & 0 & 0 & 0.1 & $3.679$ & 1 & 490 & 1 \\
	\noalign{\hrule height 1pt}
\end{tabular*}
\end{specialtable}
\begin{paracol}{2}
\switchcolumn

According to these tables, both expressions \rf{2.18} and \rf{2.20} seem preferable  for numerical calculations in the case $\tilde\lambda_{\mathrm{eff}} < 1$ since $n_{\exp}, n_{\mathrm{mix}} \ll n_{\cos}$, although  Equation~\rf{2.20} is, of course, much more complicated than  Equation~\rf{2.18}, and the computation of its every single summand takes longer. However, for $\tilde\lambda_{\mathrm{eff}} \gtrsim 1$, the Yukawa--Ewald formula \mbox{\rf{2.20}} alone becomes superior to the remaining two and the distinction grows as $\tilde\lambda_{\mathrm{eff}}$ becomes larger. 
According to Planck 2013 data \cite{36}, the lower limit on the periods of tori (in the case of chimney topology) is $\sim$20 $\rm Gpc$. Meanwhile, the current value of the effective cosmological screening length, as indicated in \cite{MaxEzgi}, is 2.6~Gpc. Thus, the region that is defined by $\tilde\lambda_{\mathrm{eff}} < 1$ depicts the observable Universe and, here, as we have just discussed, \mbox{ Equation~\rf{2.18}} 
is more convenient for numerical analysis.

\end{paracol}
\begin{specialtable}[H] 
\widetable
\caption{\label{results_table_2}Potentials $\tilde\Phi$ as well as the numbers $n_{\exp}, n_{\cos}$ and $n_{\mathrm{mix}}$ in a selection of points for  $\tilde\lambda_{\mathrm{eff}}=1$ and $\tilde\lambda_{\mathrm{eff}}=3$ in the left and right tables, respectively.}
\begin{tabular*}{\hsize}{@{}@{\extracolsep{\fill}}cccccccc|cccccccc@{}}
		\noalign{\hrule height 1pt}
		& $\bm{\tilde x}$ & $\bm{\tilde y}$ & $\bm{\tilde z}$ &$\bm{\tilde{\Phi}}$ & $\bm{n_{\exp}}$ & $\bm{n_{\cos}}$ & $\bm{n_{\mathrm{mix}}}$&& $\bm{\tilde x}$ & $\bm{\tilde y}$ & $\bm{\tilde z}$ &$\bm{\tilde{\Phi}}$ & $\bm{n_{\exp}}$ & $\bm{n_{\cos}}$ & $\bm{n_{\mathrm{mix}}}$\\	
		\noalign{\hrule height 0.5pt}
	$A_1$ & 0.5 & 0 & 0.5 & $3.783$ & 174 & 9 & 15 &$A_1$ & 0.5 & 0 & 0.5 & $15.93$ & 1418 & 7 & 6\\
	$A_2$ & 0.5 & 0 & 0.1 & $5.067$ & 163 & 229 & 15&$A_2$ & 0.5 & 0 & 0.1 & $17.60$ & 1379 & 120 & 9\\
	$A_3$ & 0.5 & 0 & 0 & $5.153$ & 163 &---& 15 &$A_3$ & 0.5 & 0 & 0 & $17.71$ & 1377 &---& 9 \\
	$B_1$ & 0.1 & 0 & 0.5 & $3.990$ & 171 & 10 & 13&$B_1$ & 0.1 & 0 & 0.5 & $16.14$ & 1411 & 8 & 6\\
	$B_2$ & 0.1 & 0 & 0.1 & $9.405$ & 133 & 164 & 11&$B_2$ & 0.1 & 0 & 0.1 & $22.00$ & 1290 & 138 & 9 \\
	$B_3$ & 0.1 & 0 & 0 & $12.34$ & 123 &--- & 10&$B_3$ & 0.1 & 0 & 0 & $24.96$ & 1242 & --- & 9\\
	$C_1$ & 0 & 0 & 0.5 & $4.014$ & 170 & 13 &13&$C_1$ & 0 & 0 & 0.5 & $16.17$ & 1410 & 8 & 7\\
	$C_2$ & 0 & 0 & 0.1 & $12.30$ & 123 & 357 &9&$C_2$ & 0 & 0 & 0.1 & $24.91$ & 1243 & 286 & 9\\
	\noalign{\hrule height 1pt}
\end{tabular*} 
\end{specialtable}
\begin{paracol}{2}
\switchcolumn

Concluding this section, we also present Figures~\ref{fig:1}--\ref{fig:4}, demonstrating the shape of the rescaled potential $\tilde\Phi$ for the same values of $\tilde\lambda_{\mathrm{eff}}$ as those picked for Tables~\ref{results_table_1} and~\ref{results_table_2}. To plot these figures (using Mathematica \cite{Math}), we use \rf{2.18} for $n \gg n_{\exp}$.

\end{paracol}


\vspace{-6pt}
\begin{paracol}{2}
\switchcolumn
\subsection{Gravitational Forces}\label{sec:4}

It is also interesting to study the forces (per unit mass) associated with the alternative forms of the gravitational potential derived in the previous section. We intend to consider the projections of these forces on the $x$- and $z$-axes.  Owing to the symmetry of the model, the $x$ and $y$ projections are similar. We calculate the gravitational forces for the same points as for the potentials and, among these, naturally, we only investigate the points at which projections on the axis of interest are nonzero. In this connection, the points $A_1,A_2,A_3,C_1,C_2$ and the points $A_3,B_3$ are omitted for the $x$- and $z$-components, respectively. The accuracy of force calculations is of the same level as that of the potentials. Here, once again, we compare the number of terms needed to achieve this accuracy, but now for three different forms of the gravitational force presentation. 
\end{paracol}
\clearpage
\vspace{-6pt}

\vspace{-6pt}

\vspace{-6pt}
\begin{paracol}{2}

\switchcolumn

\subsubsection{$x$-Component of the Gravitational Force}\label{sec:4.1}

From \rf{2.17}, \rf{2.18} and \rf{2.20}, we derive three alternative expressions for the $x$-component of the rescaled force:
\ba{4.1} 
&&\frac{\partial}{\partial \tilde x}\left(\tilde{\Phi}_{\cos}\right)=-2\pi\sum_{k_1=-\infty}^{+\infty}\sum_{k_2=-\infty}^{+\infty} 
\left(k_1^2+k_2^2+
\frac{1}{4\pi^2\tilde\lambda_{\mathrm{eff}}^2}\right)^{-1/2}\nn\\
&\times& \exp\left(-\sqrt{4\pi^2\left(k_1^2+k_2^2\right)+ \frac{1}{\tilde\lambda_{\mathrm{eff}}^2}}|\tilde z|\right) k_1\sin\left(2\pi k_1\tilde x\right)\cos\left(2\pi k_2\tilde y\right)\, ,
\ea

\end{paracol}
\nointerlineskip
\ba{4.2} 
\frac{\partial}{\partial \tilde x}\left(\tilde{\Phi}_{\exp}\right)&=&-\sum_{k_1=-\infty}^{+\infty}\sum_{k_2=-\infty}^{+\infty}\exp\left(-\frac{\sqrt{\left(\tilde x-k_1\right)^2+\left(\tilde y-k_2\right)^2+\tilde z^2}}{\tilde\lambda_{\mathrm{eff}}}\right)\nn\\
&\times&\left[\frac{\tilde x-k_1}{\left[\left(\tilde x-k_1\right)^2+\left(\tilde y-k_2\right)^2+\tilde z^2\right]^{3/2}}+\frac{\tilde x-k_1}{\tilde\lambda_{\mathrm{eff}}\left[\left(\tilde x-k_1\right)^2+\left(\tilde y-k_2\right)^2+\tilde z^2\right]}\right]\, ,\ea

\vspace{-6pt}

\begin{paracol}{2}
\switchcolumn

\end{paracol}
\nointerlineskip
\ba{4.3}
\frac{\partial}{\partial \tilde x}\left(\tilde{\Phi}_{\mathrm{mix}}\right)&=&-\frac{1}{2}\sum_{k_1=-\infty}^{+\infty}\sum_{k_2=-\infty}^{+\infty}\left[\frac{\left(\tilde x-k_1\right)D\left(\sqrt{\left(\tilde x-k_1\right)^2+\left(\tilde y-k_2\right)^2+\tilde z^2};\alpha;\tilde\lambda_{\mathrm{eff}}\right)}{\left[\left(\tilde x-k_1\right)^2+\left(\tilde y-k_2\right)^2+\tilde z^2\right]^{3/2}}\right.\nn\\
&+&\frac{\tilde x-k_1}{\left(\tilde x-k_1\right)^2+\left(\tilde y-k_2\right)^2+\tilde z^2}\cdot C_-\exp\left(-\frac{\sqrt{\left(\tilde x-k_1\right)^2+\left(\tilde y-k_2\right)^2+\tilde z^2}}{\tilde\lambda_{\mathrm{eff}}}\right)\,\nn\\
&+&\frac{\tilde x-k_1}{\left(\tilde x-k_1\right)^2+\left(\tilde y-k_2\right)^2+\tilde z^2}\cdot C_+\exp\left(\frac{\sqrt{\left(\tilde x-k_1\right)^2+\left(\tilde y-k_2\right)^2+\tilde z^2}}{\tilde\lambda_{\mathrm{eff}}}\right)\nn\\
&+&\left.
4\pi^2  k_1\sin\left[2\pi\left(k_1\tilde x+k_2\tilde y\right)\right]\frac{F\left(\sqrt{4\pi^2\left(k_1^2+k_2^2\right)+\tilde\lambda^{-2}_{\mathrm{eff}}};\tilde z;\alpha\right)}{\sqrt{4\pi^2\left(k_1^2+k_2^2\right)+\tilde\lambda^{-2}_{\mathrm{eff}}}}\right]\, ,
\ea
\begin{paracol}{2}
\switchcolumn
\noindent where
\ba{4.4}
C_{\mp}&=&C_{\mp}\left(\sqrt{\left(\tilde x-k_1\right)^2+\left(\tilde y-k_2\right)^2+\tilde z^2};\alpha;\tilde\lambda_{\mathrm{eff}}\right)\,\nn\\
&\equiv&\frac{2\alpha}{\sqrt{\pi}}\exp\left[-\left(\alpha\sqrt{\left(\tilde x-k_1\right)^2+\left(\tilde y-k_2\right)^2+\tilde z^2}\mp\frac{1}{2\alpha\tilde\lambda_{\mathrm{eff}}}\right)^2\right]\,\nn\\
&\pm&\frac{1}{\tilde\lambda_{\mathrm{eff}}}\mathrm{erfc}\left(\alpha\sqrt{\left(\tilde x-k_1\right)^2+\left(\tilde y-k_2\right)^2+\tilde z^2}\mp\frac{1}{2\alpha\tilde\lambda_{\mathrm{eff}}}\right)\, .
\ea

We present the results of the calculations that were performed in Mathematica \cite{Math} in \mbox{Tables~\ref{results_table_3}} and~\ref{results_table_4} for $\tilde\lambda_{\mathrm{eff}}=0.01, 0.1$ and $\tilde\lambda_{\mathrm{eff}}=1,3$, respectively. As in the case for the gravitational potential, a straightforward analysis shows that Formulas \rf{4.2} and \rf{4.3} that are related to the Yukawa and Yukawa--Ewald potentials, respectively, are preferable \mbox{over~\rf{4.1}} for the physically significant case $\tilde\lambda_{\mathrm{eff}} < 1$ (although the structure of the \mbox{expression \mbox{\rf{4.3}}} is again much more complicated compared to \rf{4.2}). Meanwhile, when  $\tilde\lambda_{\mathrm{eff}} \gtrsim 1$, the Yukawa--Ewald force becomes superior. In both tables,  Equation~\rf{4.2} was employed for $n\gg n_{\exp}$ while computing the values of the rescaled $x$-component $\tilde{\Phi}_x$.
\end{paracol}
\begin{specialtable}[H] 
\widetable
\caption{\label{results_table_3}Numerical values of the $x$-component of the rescaled force $\tilde{\Phi}_x\equiv \partial \tilde\Phi/\partial\tilde x$ as well as the numbers $n_{\exp}, n_{\cos}$ and $n_{\mathrm{mix}}$ for points $B_1, B_2$ and $B_3$ for $\tilde\lambda_{\mathrm{eff}}=0.01$ and $\tilde\lambda_{\mathrm{eff}}=0.1$ in the left and right tables, respectively.}
	
\begin{tabular*}{\hsize}{@{}@{\extracolsep{\fill}}cccccccc|cccccccc@{}}
		\noalign{\hrule height 1pt}
		& $\bm{\tilde x}$ & $\bm{\tilde y}$ & $\bm{\tilde z}$ &$\bm{\tilde{\Phi}_x}$ & $\bm{n_{\exp}}$ & $\bm{n_{\cos}}$ & $\bm{n_{\mathrm{mix}}}$&& $\bm{\tilde x}$ & $\bm{\tilde y}$ & $\bm{\tilde z}$ &$\bm{\tilde{\Phi}_x}$ & $\bm{n_{\exp}}$ & $\bm{n_{\cos}}$ & $\bm{n_{\mathrm{mix}}}$\\
		
		\noalign{\hrule height 0.5pt}

		$B_1$ & 0.1 & 0 & 0.5 & $-2.810\times10^{-21}$ & 1 & 263& 1&$B_1$ & 0.1 & 0 & 0.5 & $-2.783\times10^{-2}$ & 5 & 54 &5 \\

		$B_2$ & 0.1 & 0 & 0.1 & $-3.862\times10^{-4}$ & 1 & 2448 & 1& $B_2$ & 0.1 & 0 & 0.1 & $-20.75$ & 1 & 592 & 1\\

		$B_3$ & 0.1 & 0 & 0 & $-4.994\times10^{-2}$ & 1 & --- & 1& $B_3$ & 0.1 & 0 &0 & $-73.57$ & 1 & --- & 1\\
		
		\noalign{\hrule height 1pt}		
	\end{tabular*}
	\end{specialtable}
\vspace{-6pt}

\begin{specialtable}[H] 
\widetable
\caption{\label{results_table_4}Numerical values of the $x$-component of the rescaled force $\tilde{\Phi}_x\equiv \partial \tilde\Phi/\partial\tilde x$ as well as the numbers $n_{\exp}, n_{\cos}$ and $n_{\mathrm{mix}}$ for points $B_1, B_2$ and $B_3$ for $\tilde\lambda_{\mathrm{eff}}=1$ and $\tilde\lambda_{\mathrm{eff}}=3$ in the left and right tables, respectively.}
\begin{tabular*}{\hsize}{@{}@{\extracolsep{\fill}}cccccccc|cccccccc@{}}
		\noalign{\hrule height 1pt}
		& $\bm{\tilde x}$ & $\bm{\tilde y}$ & $\bm{\tilde z}$ &$\bm{\tilde{\Phi}_x}$ & $\bm{n_{\exp}}$ & $\bm{n_{\cos}}$ & $\bm{n_{\mathrm{mix}}}$&& $\bm{\tilde x}$ & $\bm{\tilde y}$ & $\bm{\tilde z}$ &$\bm{\tilde{\Phi}_x}$ & $\bm{n_{\exp}}$ & $\bm{n_{\cos}}$ & $\bm{n_{\mathrm{mix}}}$\\	
		\noalign{\hrule height 0.5pt}
		$B_1$ & 0.1 & 0 & 0.5 & $-4.730\times10^{-1}$ & 130 & 38 & 21&$B_1$ & 0.1 & 0 & 0.5 & $-4.920\times10^{-1}$ & 862 & 38 & 21\\ 
		$B_2$ & 0.1 & 0 & 0.1 & $-34.65$ & 20 & 553 & 9&$B_2$ & 0.1 & 0 & 0.1 & $-34.88$ & 77 & 552 & 13\\ 
		$B_3$ & 0.1 & 0 & 0 & $-99.14$ & 19 & ---  & 8&$B_3$ & 0.1 & 0 & 0 & $-99.49$ & 34 & --- & 9\\		
		\noalign{\hrule height 1pt}
		\end{tabular*}
	\end{specialtable}
	
\begin{figure}[H]
\widefigure
\resizebox{0.46\textwidth}{!}{\includegraphics{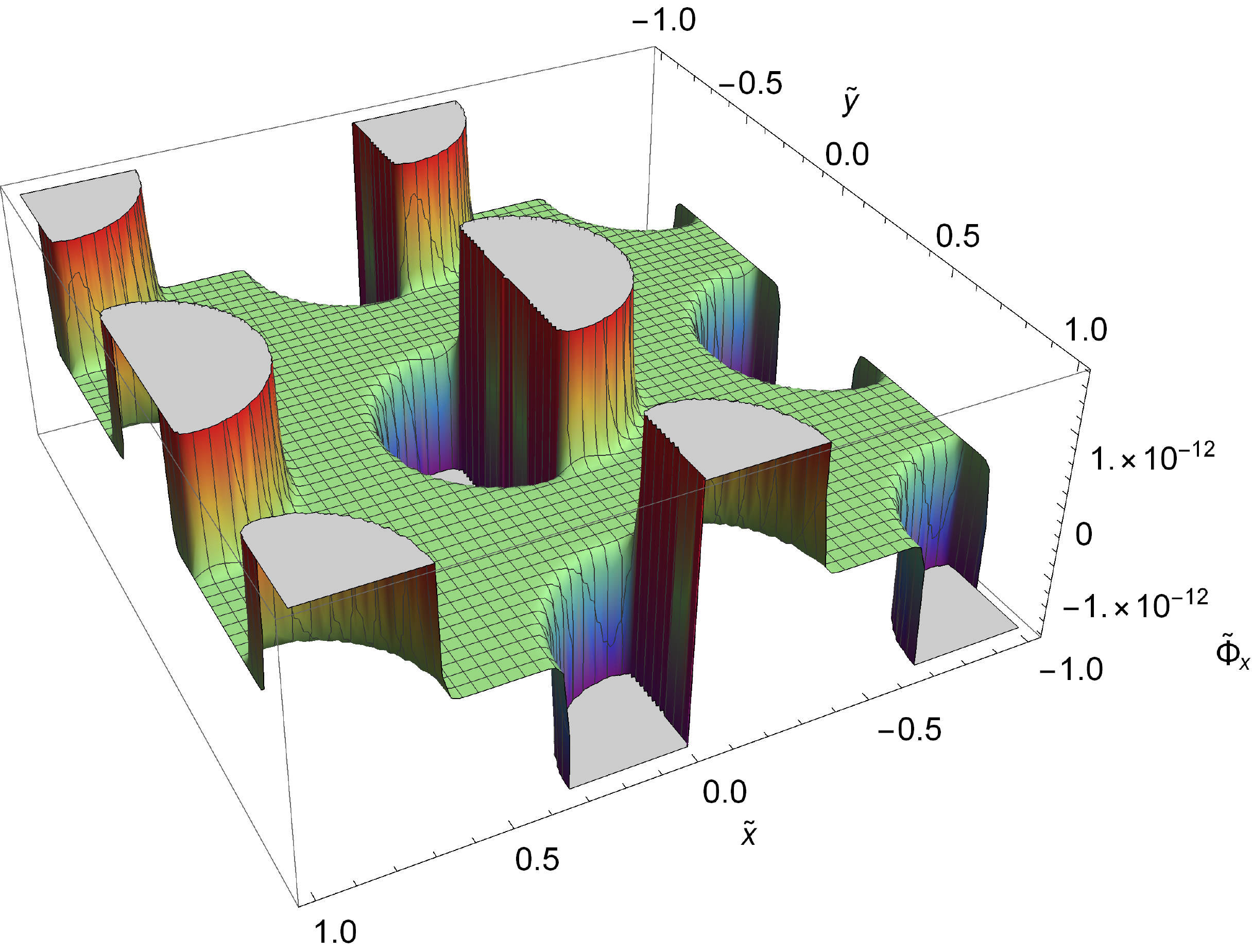}}\quad\quad
	\resizebox{0.46\textwidth}{!}{\includegraphics{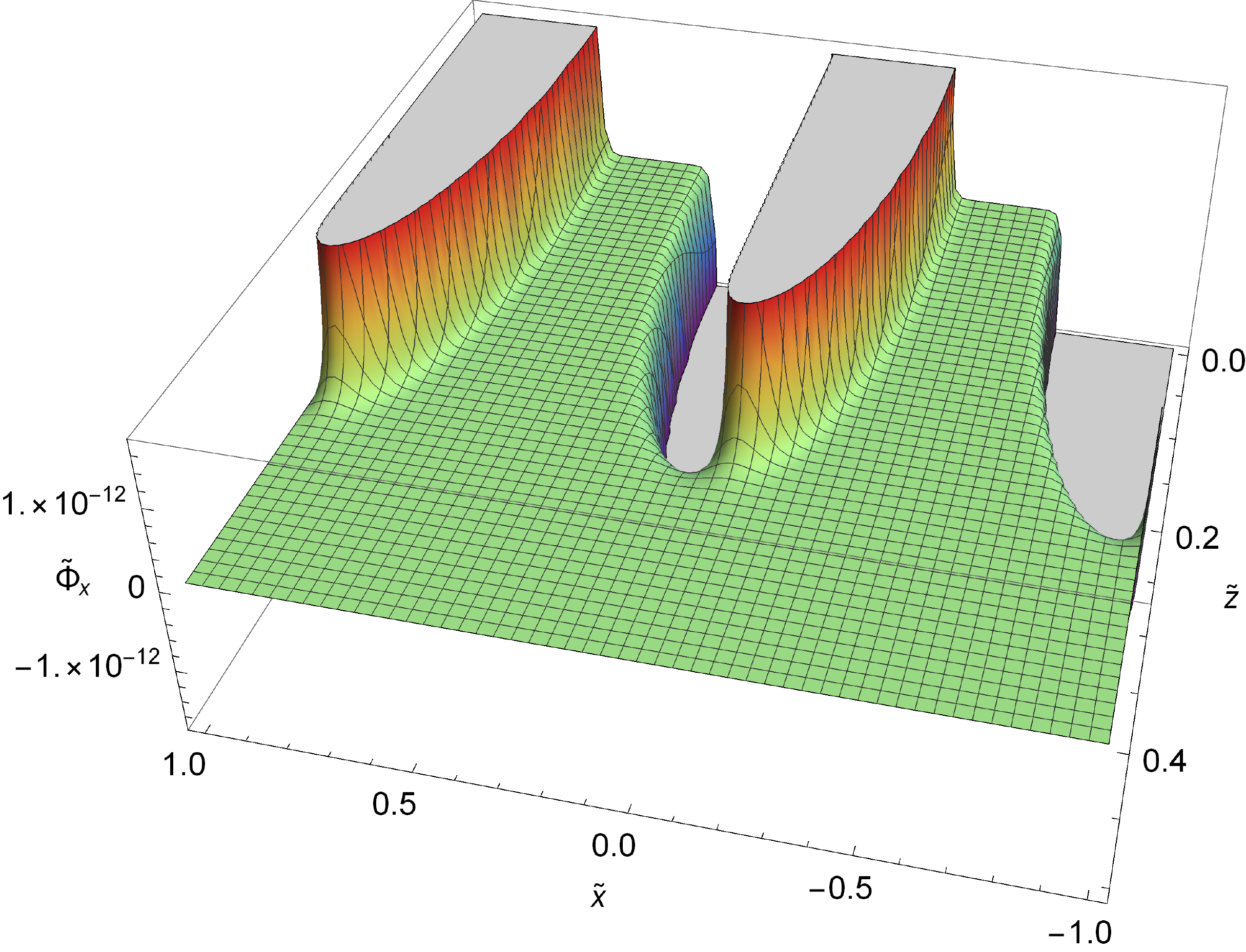}}
\caption{$x$-component of the rescaled force $\tilde{\Phi}_x$ for $\tilde\lambda_{\mathrm{eff}} = 0.01$ for the sections $z=0$ (left panel) and $y=0$ (right panel).\label{fig:5}}
\end{figure}

\begin{figure}[H]
\widefigure
\resizebox{0.46\textwidth}{!}{\includegraphics{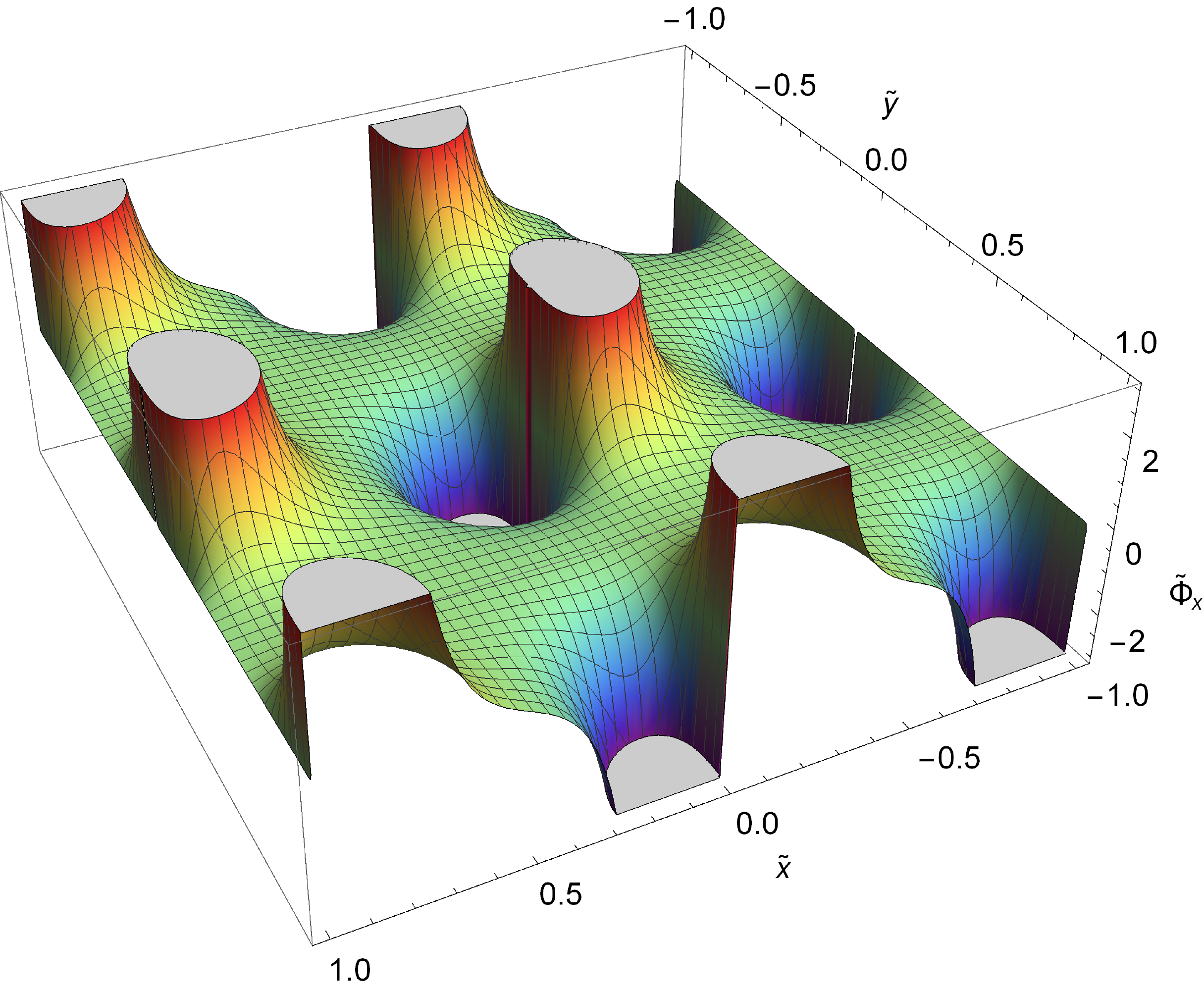}}\quad\quad
	\resizebox{0.46\textwidth}{!}{\includegraphics{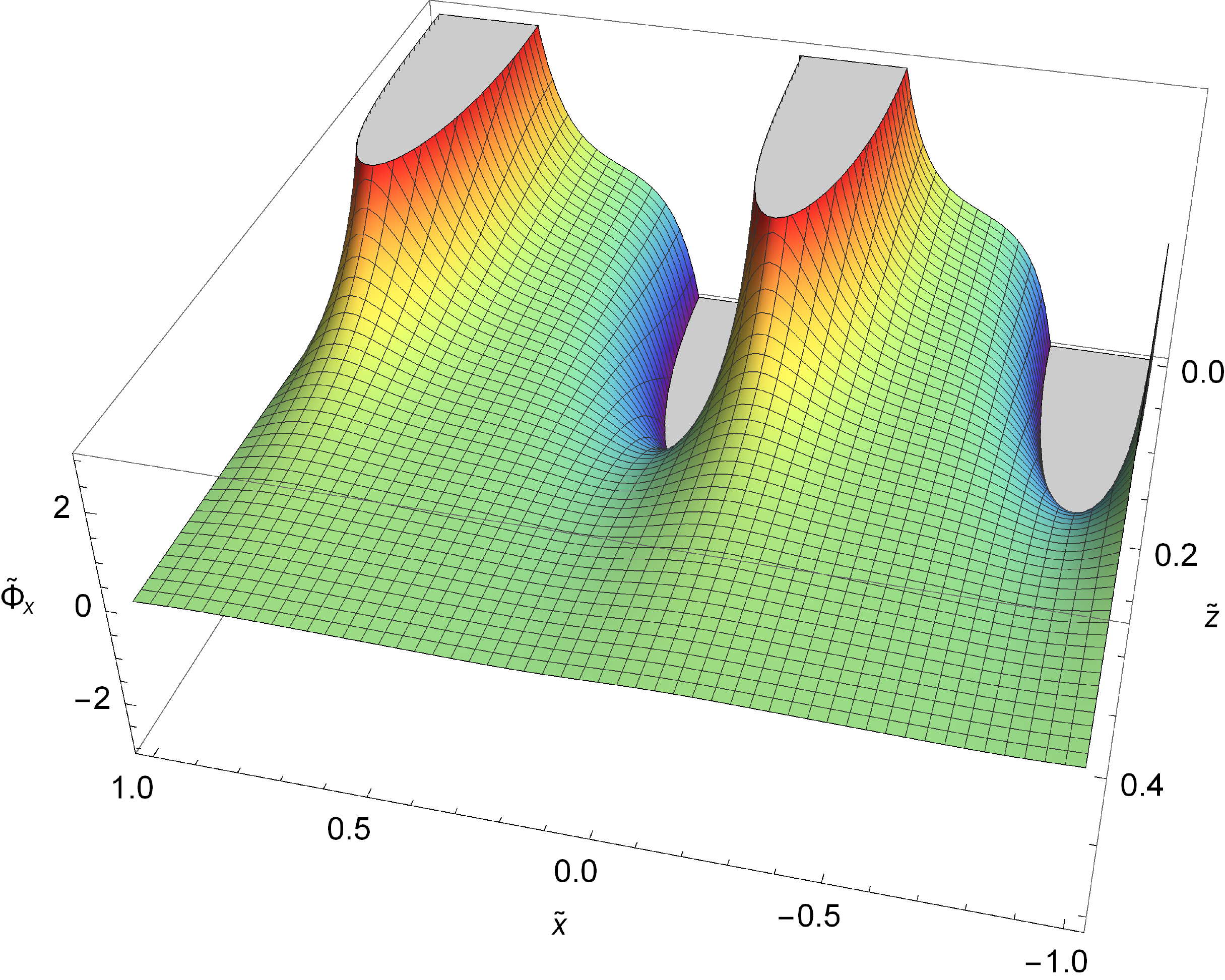}}
\caption{$x$-component of the rescaled force $\tilde{\Phi}_x$ for $\tilde\lambda_{\mathrm{eff}} = 0.1$ for the sections $z=0$ (left panel) and $y=0$ (right panel).\label{fig:6}}
\end{figure}

\begin{figure}[H]
\widefigure
\resizebox{0.47\textwidth}{!}{\includegraphics{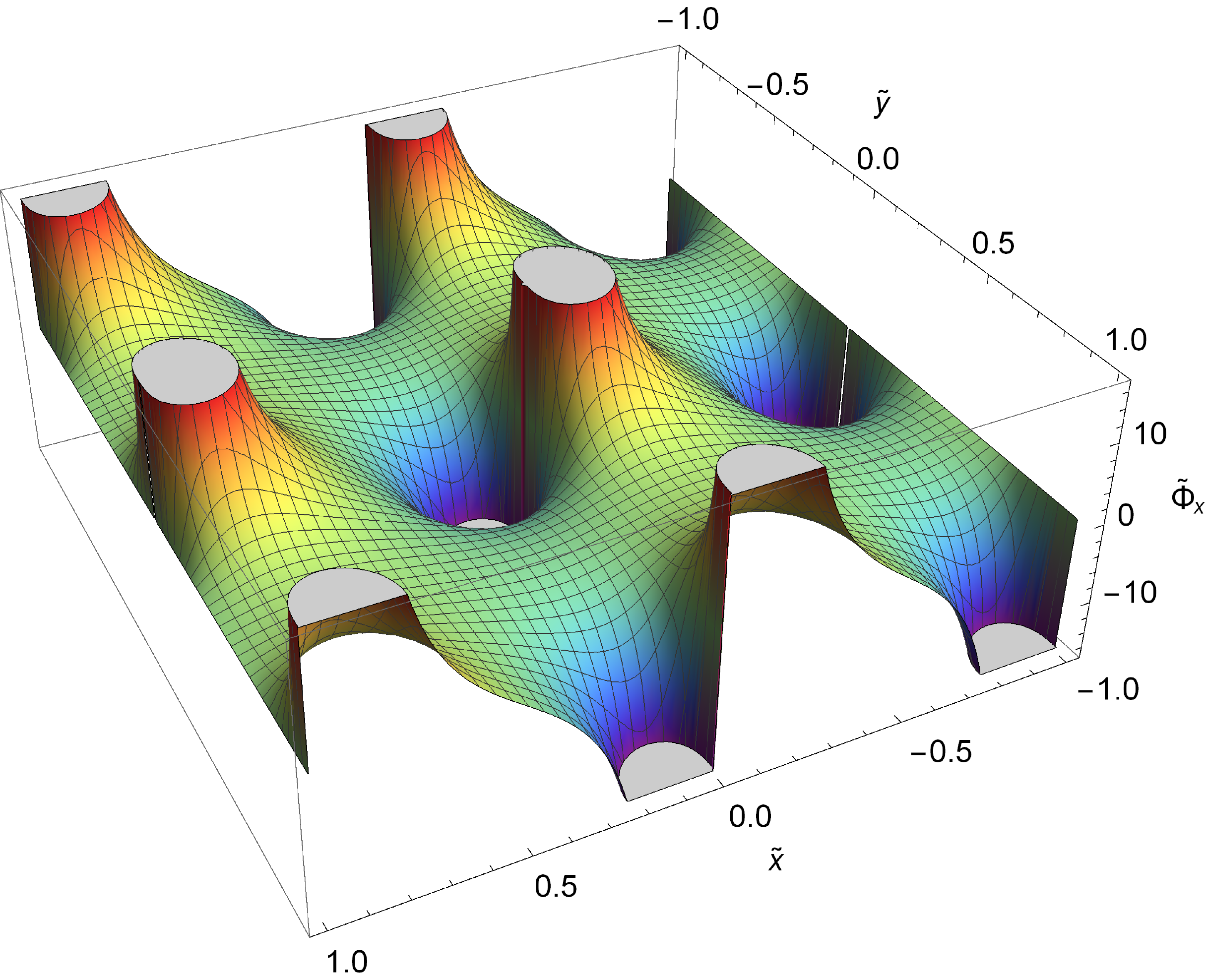}}\quad\quad
	\resizebox{0.47\textwidth}{!}{\includegraphics{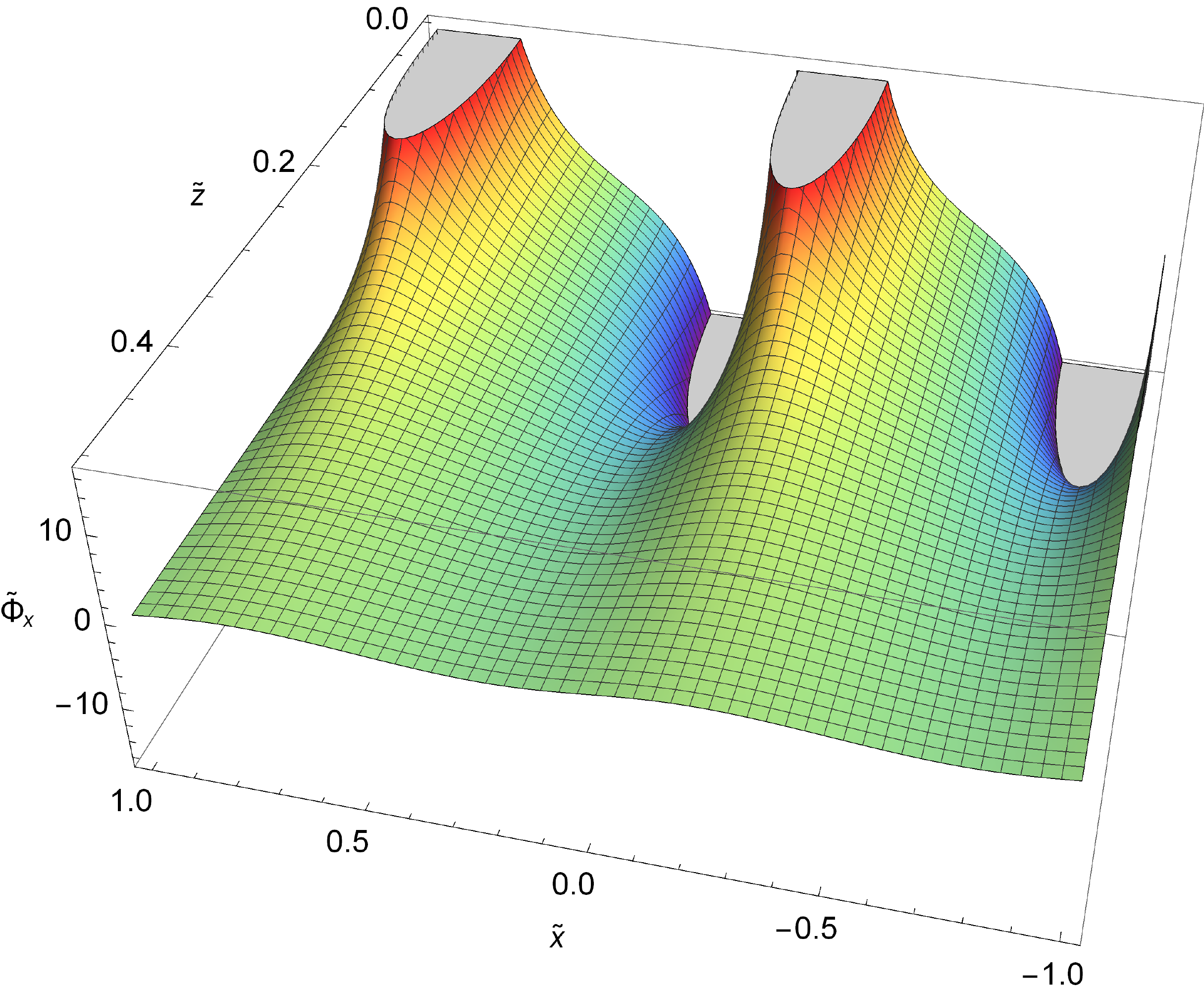}}\caption{$x$-component of the rescaled force $\tilde{\Phi}_x$ for $\tilde\lambda_{\mathrm{eff}} = 1$ for the sections $z=0$ (left panel) and $y=0$ (right panel).\label{fig:7}}
\end{figure}

\begin{figure}[H]
\widefigure
\resizebox{0.47\textwidth}{!}{\includegraphics{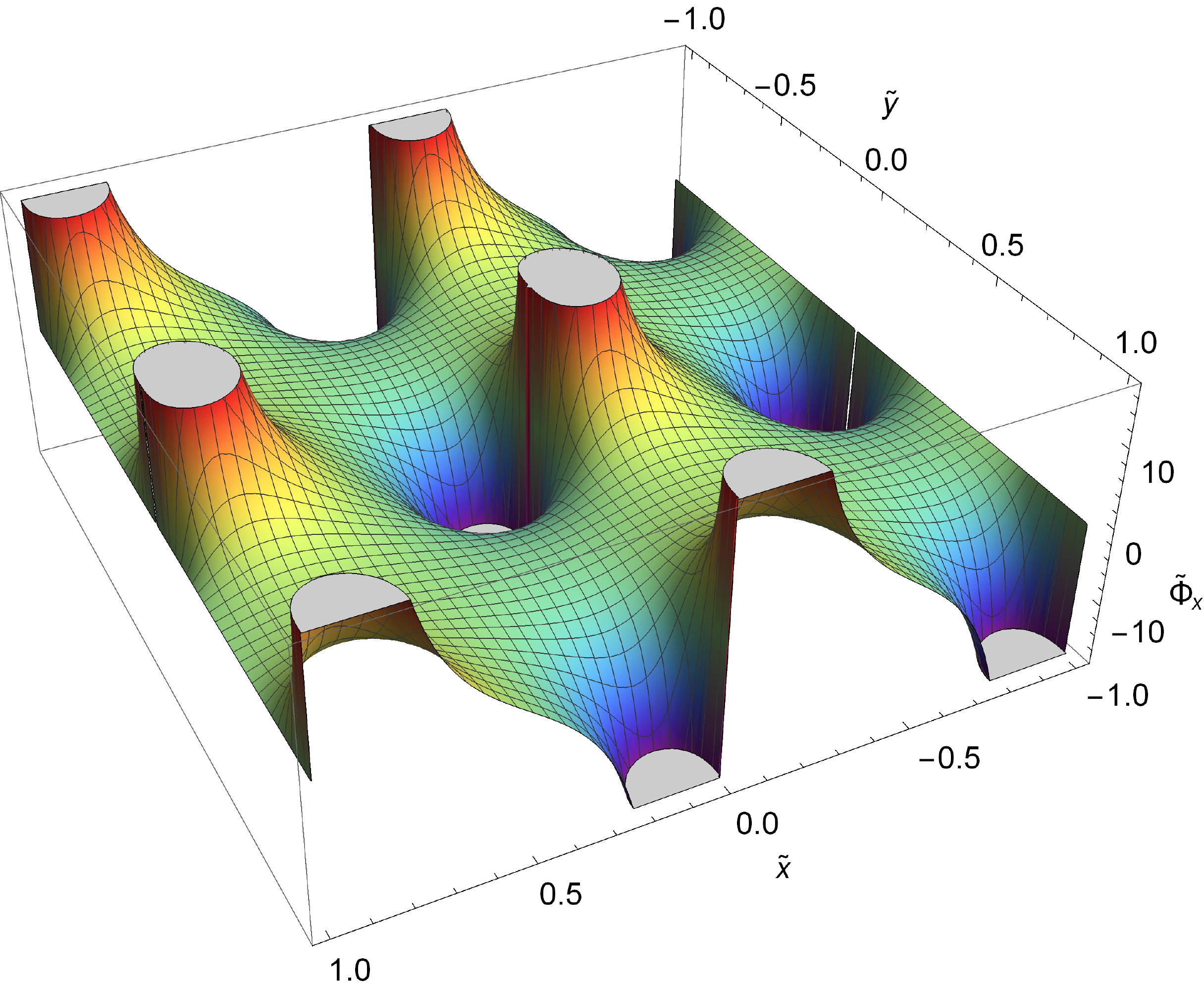}}\quad\quad
	\resizebox{0.47\textwidth}{!}{\includegraphics{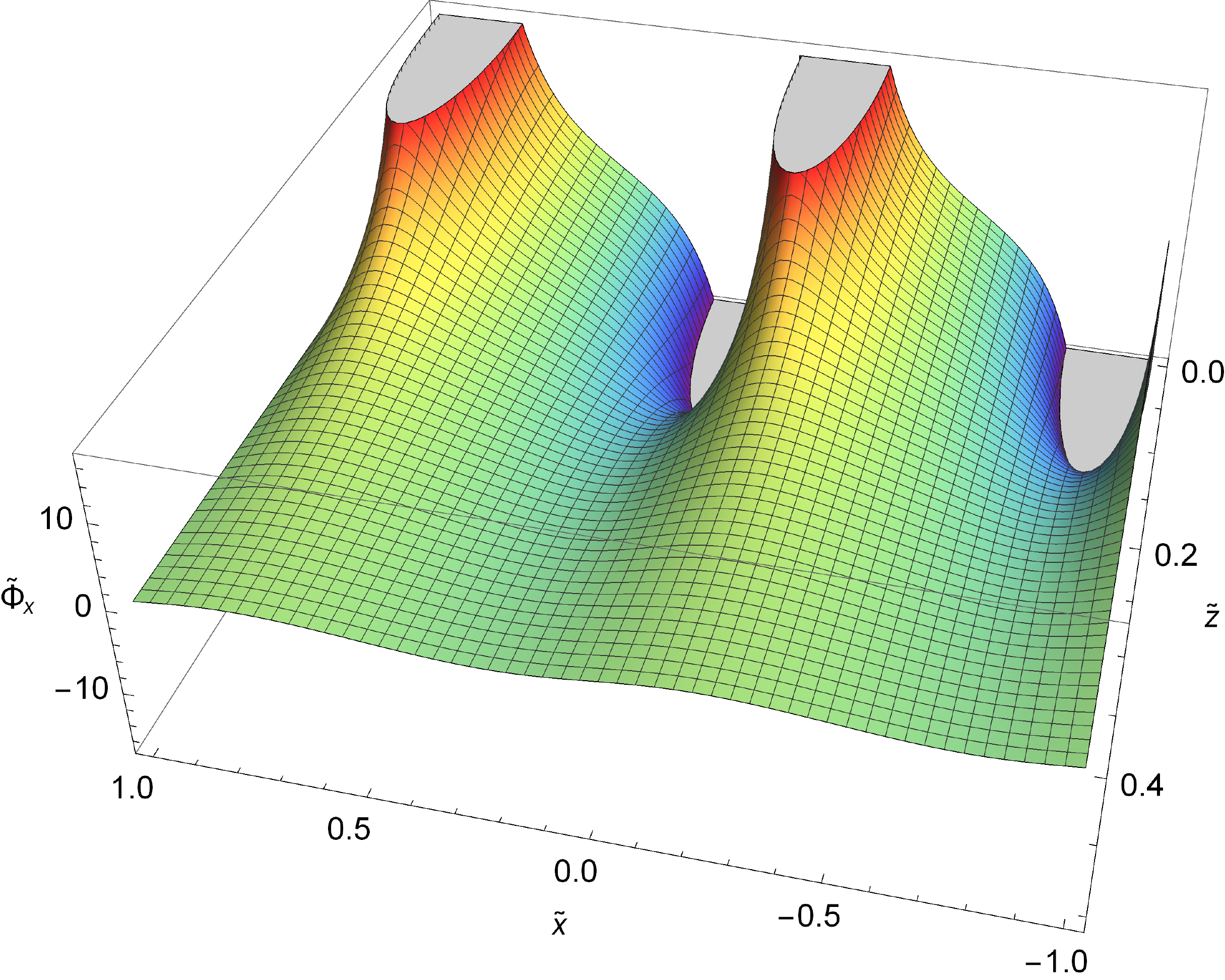}}\caption{$x$-component of the rescaled force $\tilde{\Phi}_x$ for $\tilde\lambda_{\mathrm{eff}} = 3$ for the sections $z=0$ (left panel) and $y=0$ (right panel).\label{fig:8}}
\end{figure}

\begin{paracol}{2}
\switchcolumn

Additionally, we present Figures~\ref{fig:5}--\ref{fig:8}, demonstrating the $x$-component of the rescaled force $\tilde{\Phi}_x$ for the same values of $\tilde\lambda_{\mathrm{eff}}$ as those picked for Tables~\ref{results_table_3} and~\ref{results_table_4}. To plot these figures (using Mathematica \cite{Math}), we employ the Formula \rf{4.2} for $n \gg n_{\exp}$. 
\clearpage
\end{paracol}

\vspace{-6pt}

\vspace{-6pt}

\vspace{-6pt}

\begin{figure}[H]
\widefigure
\resizebox{0.47\textwidth}{!}{\includegraphics{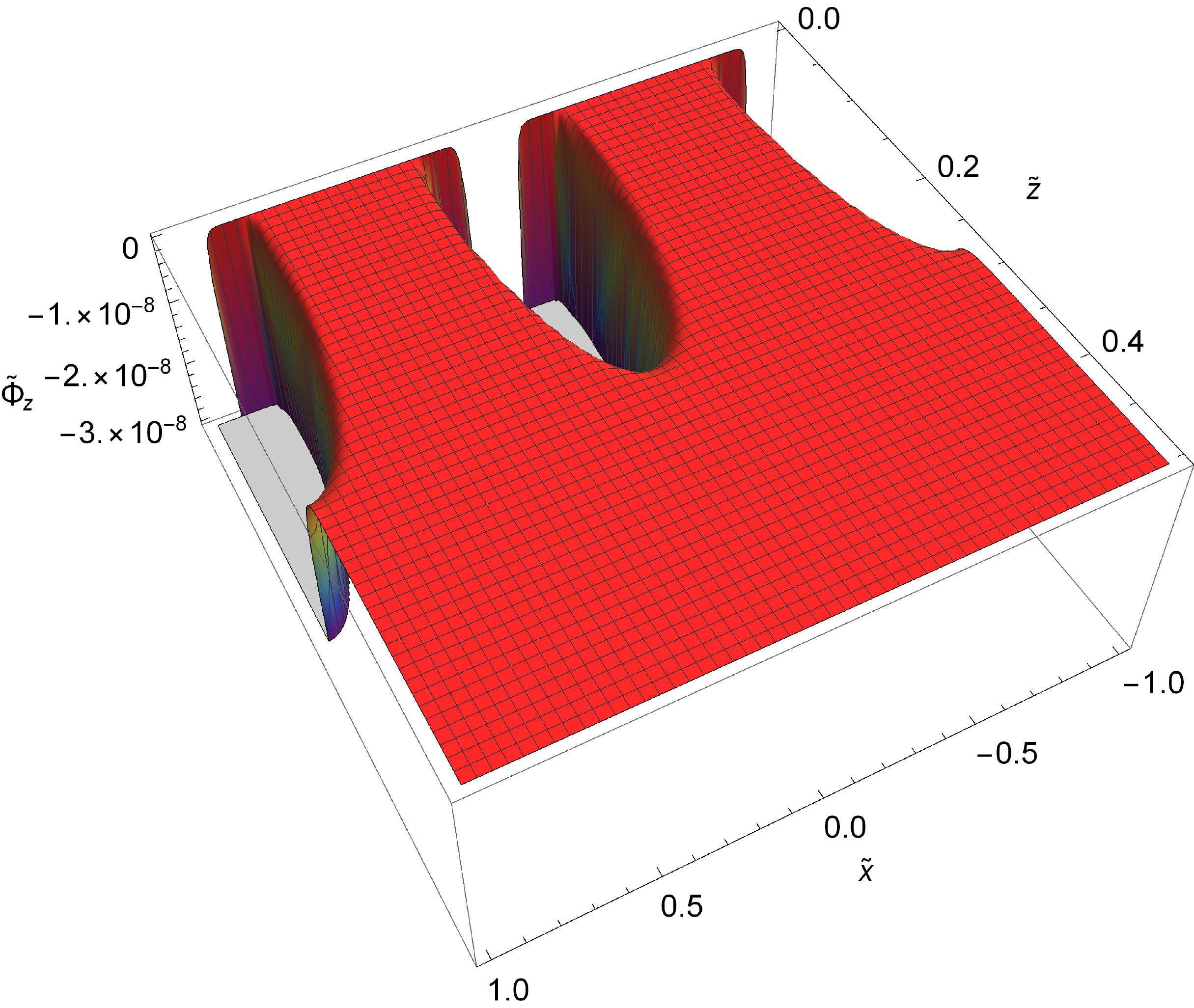}}\quad\quad
	\resizebox{0.47\textwidth}{!}{\includegraphics{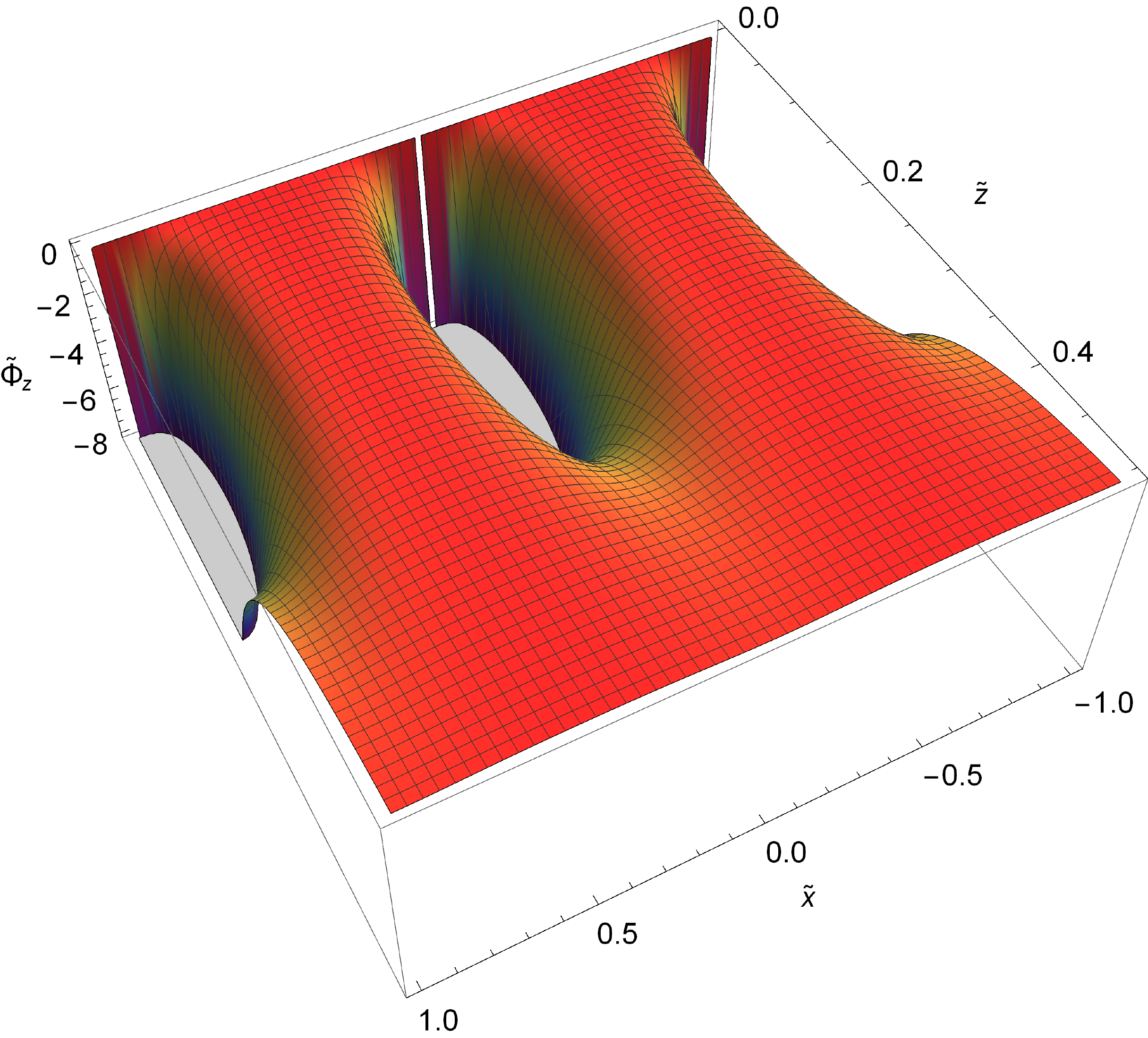}}
\caption{$z$-component of the rescaled force $\tilde{\Phi}_{z}$ for $\tilde\lambda_{\mathrm{eff}} = 0.01$ and $\tilde\lambda_{\mathrm{eff}} = 0.1$ (left and right panels, respectively).\label{fig:9}}
\end{figure}

\begin{figure}[H]	
\widefigure
\resizebox{0.47\textwidth}{!}{\includegraphics{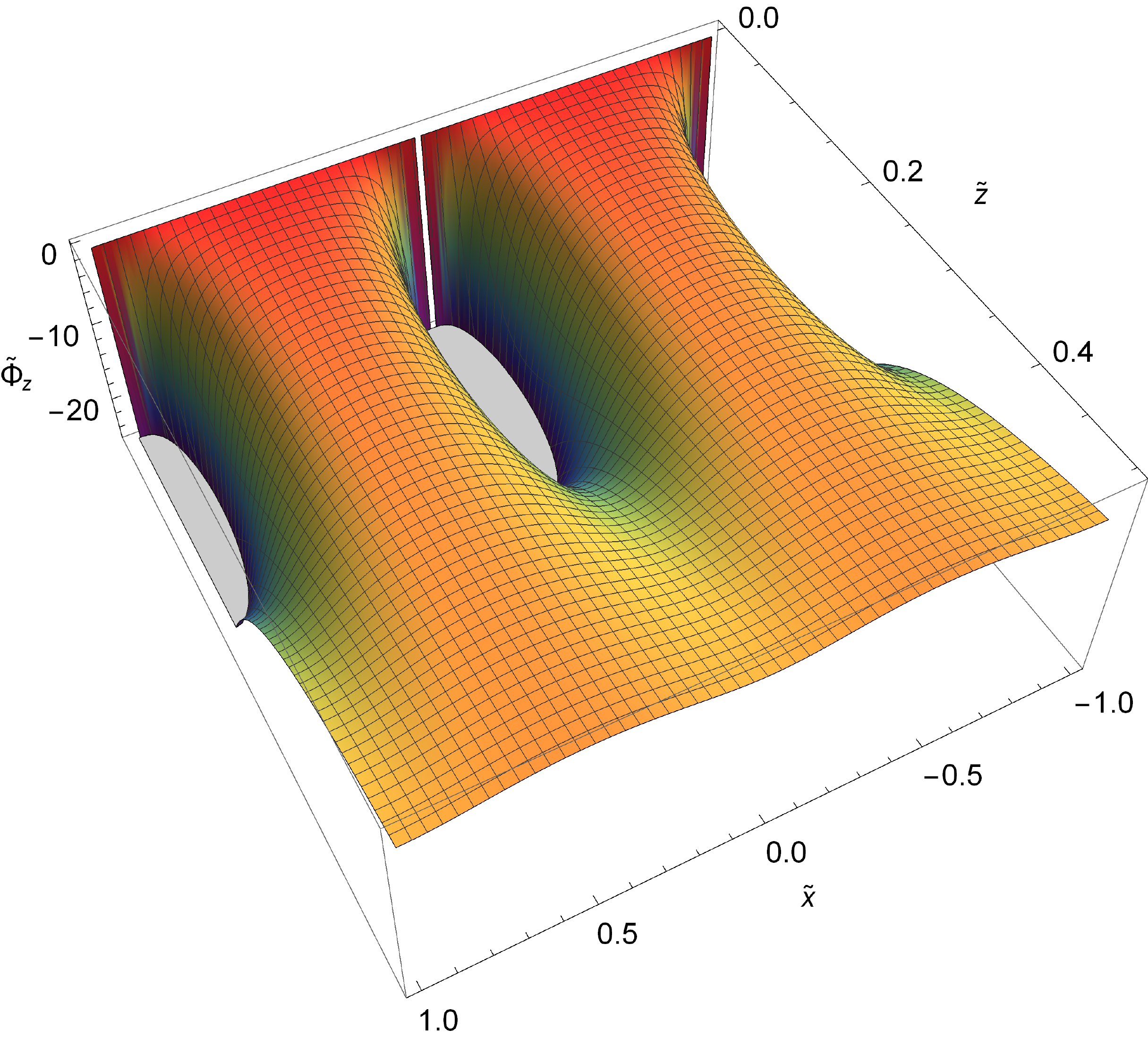}}\quad\quad
	\resizebox{0.47\textwidth}{!}{\includegraphics{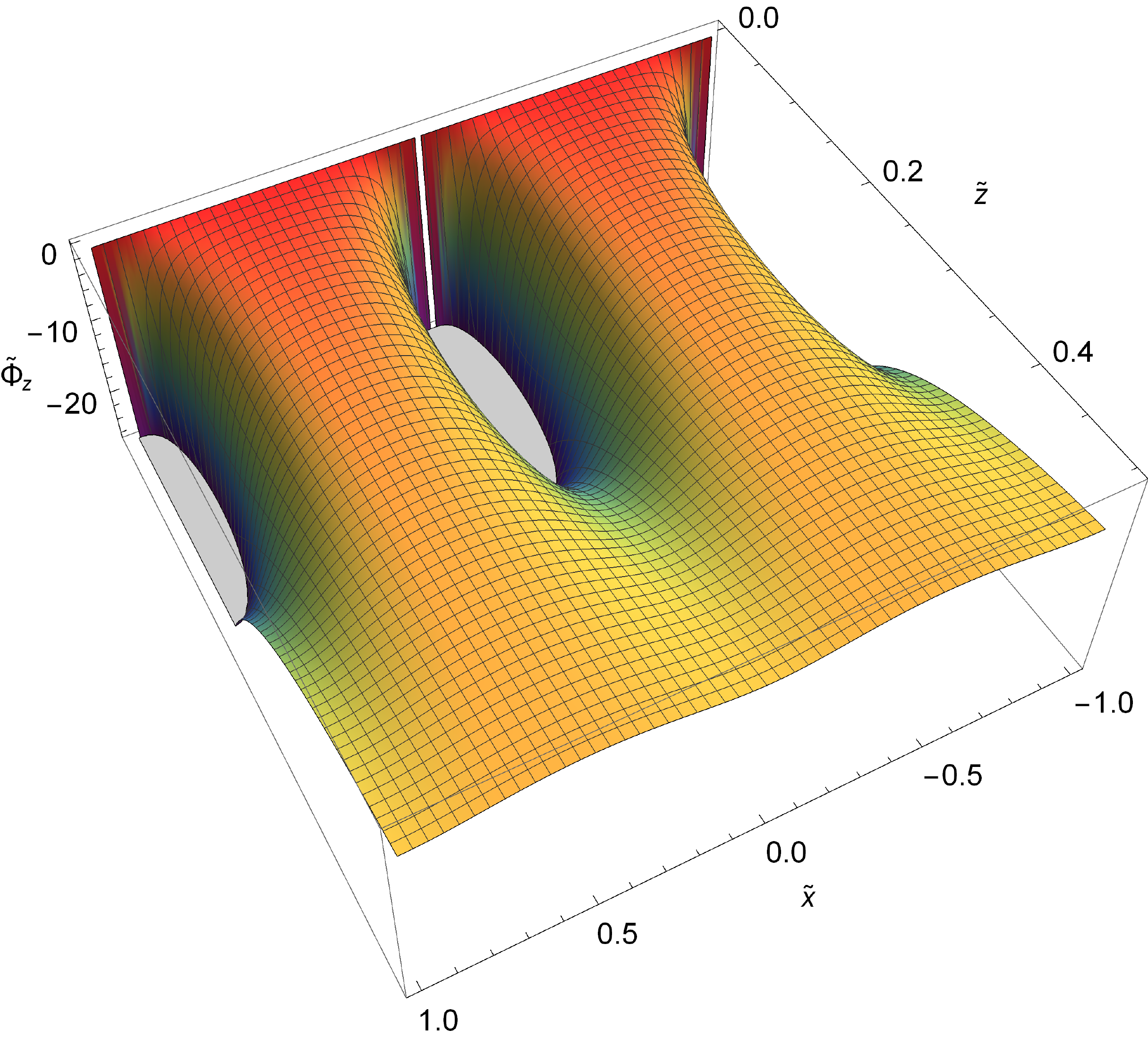}}
\caption{$z$-component of the rescaled force $\tilde{\Phi}_{z}$ for $\tilde\lambda_{\mathrm{eff}} = 1$ and $\tilde\lambda_{\mathrm{eff}} = 3$ (left and right panels, respectively).\label{fig:10}}
\end{figure}

\vspace{-12pt} 
\begin{paracol}{2}
\switchcolumn
\subsubsection{$z$-Component of the Gravitational Force}\label{sec:4.2}

For the $z$-component of the rescaled force, three
alternative formulas are:
\ba{4.5} \frac{\partial}{\partial \tilde z}\left(\tilde{\Phi}_{\cos}\right)&=&-2\pi\sum_{k_1=-\infty}^{+\infty}\sum_{k_2=-\infty}^{+\infty} \exp\left(-\sqrt{4\pi^2\left(k_1^2+k_2^2\right)+ \frac{1}{\tilde\lambda_{\mathrm{eff}}^2}}\tilde z\right)\,\nn\\
&\times&\cos\left(2\pi k_1\tilde x\right)\cos\left(2\pi k_2\tilde y\right)\, ,\ea

where, for simplicity, $\tilde z>0$,
\end{paracol}
\nointerlineskip
\ba{4.6} 
\frac{\partial}{\partial \tilde z}\left(\tilde{\Phi}_{\exp}\right)&=&-\sum_{k_1=-\infty}^{+\infty}\sum_{k_2=-\infty}^{+\infty}\exp\left(-\frac{\sqrt{\left(\tilde x-k_1\right)^2+\left(\tilde y-k_2\right)^2+\tilde z^2}}{\tilde\lambda_{\mathrm{eff}}}\right)\nn\\
&\times&\left[\frac{\tilde z}{\left[\left(\tilde x-k_1\right)^2+\left(\tilde y-k_2\right)^2+\tilde z^2\right]^{3/2}}+\frac{\tilde z}{\tilde\lambda_{\mathrm{eff}}\left[\left(\tilde x-k_1\right)^2+\left(\tilde y-k_2\right)^2+\tilde z^2\right]}\right]\, \ea
\begin{paracol}{2}
\switchcolumn

\noindent and
\end{paracol}
\nointerlineskip
\ba{4.7}
&&\frac{\partial}{\partial \tilde z}\left(\tilde{\Phi}_{\mathrm{mix}}\right)=\,\nn\\
&-&\frac{1}{2}\sum_{k_1=-\infty}^{+\infty}\sum_{k_2=-\infty}^{+\infty}\left[\frac{\tilde z D\left(\sqrt{\left(\tilde x-k_1\right)^2+\left(\tilde y-k_2\right)^2+\tilde z^2};\alpha;\tilde\lambda_{\mathrm{eff}}\right)}{\left[\left(\tilde x-k_1\right)^2+\left(\tilde y-k_2\right)^2+\tilde z^2\right]^{3/2}}\right.\,\nn\\
&+&\frac{\tilde z}{\left(\tilde x-k_1\right)^2+\left(\tilde y-k_2\right)^2+\tilde z^2}\cdot C_-\exp\left(-\frac{\sqrt{\left(\tilde x-k_1\right)^2+\left(\tilde y-k_2\right)^2+\tilde z^2}}{\tilde\lambda_{\mathrm{eff}}}\right)\,\nn\\
&+&\frac{\tilde z}{\left(\tilde x-k_1\right)^2+\left(\tilde y-k_2\right)^2+\tilde z^2}\cdot C_+\exp\left(\frac{\sqrt{\left(\tilde x-k_1\right)^2+\left(\tilde y-k_2\right)^2+\tilde z^2}}{\tilde\lambda_{\mathrm{eff}}}\right)\,\nn\\
&-&2\pi\cos\left[2\pi \left(k_1\tilde x+k_2\tilde y\right)\right]\frac{1}{\sqrt{4\pi^2\left(k_1^2+k_2^2\right)+\tilde\lambda^{-2}_{\mathrm{eff}}}} \nn\\
&\times&\left.\left[F_-\exp\left(-z\sqrt{4\pi^2\left(k_1^2+k_2^2\right)+\frac{1}{\tilde\lambda^{2}_{\mathrm{eff}}}}\right)+F_+\exp\left(z\sqrt{4\pi^2\left(k_1^2+k_2^2\right)+\frac{1}{\tilde\lambda^{2}_{\mathrm{eff}}}}\right)\right]\right] \, ,\ea
\begin{paracol}{2}
\switchcolumn

\noindent where
\ba{20} F_{\mp}&=&F_{\mp}\left(\sqrt{4\pi^2\left(k_1^2+k_2^2\right)+\tilde\lambda^{-2}_{\mathrm{eff}}};\tilde z;\alpha\right)\equiv\,\nn\\
&\pm&\frac{2\alpha}{\sqrt{\pi}}\exp\left[-\left(\frac{\sqrt{4\pi^2\left(k_1^2+k_2^2\right)+\tilde\lambda^{-2}_{\mathrm{eff}}}}{2\alpha}\mp\alpha \tilde z\right)^2\right]\nn\\
&\mp&\sqrt{4\pi^2\left(k_1^2+k_2^2\right)+\tilde\lambda^{-2}_{\mathrm{eff}}}\mathrm{erfc}\left(\frac{\sqrt{4\pi^2\left(k_1^2+k_2^2\right)+\tilde\lambda^{-2}_{\mathrm{eff}}}}{2\alpha}\mp\alpha \tilde z\right) \, ,\ea
and $C_{\mp}$ are given by \rf{4.4}.

Now, we employ these formulas to calculate the nonzero $z$-components of the gravitational force at the previously selected set of points and, again, for the desired precision. The results that were obtained in Mathematica \cite{Math} are presented in Tables~\ref{results_table_5} and \ref{results_table_6}, which show that while $\tilde\lambda_{\mathrm{eff}}<1$, depicting well the observational restrictions, two \mbox{Formulas \rf{4.6} and \rf{4.7}} are favorable (as before, the latter is much more cumbersome). On the other hand, for $\tilde\lambda_{\mathrm{eff}}\gtrsim1$, the Yukawa--Ewald Formula \rf{4.7} gives the best results.
Herein, the quantity $\tilde{\Phi}_{z}$ stands for the $z$-component of the rescaled force, calculated from  Equation~\rf{4.6} for $n\gg n_{\exp}$.  We depict the behavior of this component in Figures~\ref{fig:9} and \ref{fig:10} for the section $y=0$. Obviously, the projection of the gravitational force on the $z$-axis is absent for the section $z=0$ due to the symmetry of the model.
\end{paracol}
\begin{specialtable}[H] 
\widetable
\caption{\label{results_table_5}Numerical values of the $z$-component of the rescaled force $\tilde{\Phi}_{z}\equiv \partial \tilde\Phi/\partial\tilde z$ as well as the numbers $n_{\exp}, n_{\cos}$ and $n_{\mathrm{mix}}$ for points $A_1, A_2, B_1, B_2, C_1$ and $C_2$ for $\tilde\lambda_{\mathrm{eff}}=0.01$ and $\tilde\lambda_{\mathrm{eff}}=0.1$ in the left and right tables, respectively.}
\begin{tabular*}{\hsize}{@{}@{\extracolsep{\fill}}cccccccc|cccccccc@{}}
		\noalign{\hrule height 1pt}
		& $\bm{\tilde x}$ & $\bm{\tilde y}$ & $\bm{\tilde z}$ &$\bm{\tilde{\Phi}_z}$ & $\bm{n_{\exp}}$ & $\bm{n_{\cos}}$ & $\bm{n_{\mathrm{mix}}}$&& $\bm{\tilde x}$ & $\bm{\tilde y}$ & $\bm{\tilde z}$ &$\bm{\tilde{\Phi}_z}$ & $\bm{n_{\exp}}$ & $\bm{n_{\cos}}$ & $\bm{n_{\mathrm{mix}}}$\\
	\noalign{\hrule height 0.5pt}	
		$A_1$ & 0.5 & 0 & 0.5 & $-3.962\times10^{-29}$ & 2 & 1070 & 2&$A_1$ & 0.5 & 0 & 0.5 &  $-1.946\times10^{-2}$ & 6 & 47 & 6\\		
		$A_2$ & 0.5 & 0 & 0.1 &  $-5.620\times10^{-21}$ & 2 & --- & 2&	$A_2$ & 0.5 & 0 & 0.1 &  $-5.620\times10^{-2}$ & 2 & 1647 & 2\\		
		$B_1$ & 0.1 & 0 & 0.5 &  $-1.405\times10^{-20}$ & 1 & 187 & 1&$B_1$ & 0.1 & 0 & 0.5 &  $-1.407\times10^{-1}$ & 2 & 33 & 2\\		
		$B_2$ & 0.1 & 0 & 0.1 & $-3.862\times10^{-4}$ & 1 & 2228 & 1&$B_2$ & 0.1 & 0 & 0.1 &  $-20.75$ & 1 & 649 & 1\\		
		$C_1$ & 0 & 0 & 0.5 &  $-3.935\times10^{-20}$ & 1 & 240 & 1& $C_1$ & 0 & 0 & 0.5 &  $-1.620\times10^{-1}$ & 3 & 44 & 3\\		
		$C_2$ & 0 & 0 & 0.1 &  $-4.994\times10^{-2}$ & 1 & 1620 & 1& $C_2$ & 0 & 0 & 0.1 &  $-73.58$ & 1 & 722 & 1\\		
		\noalign{\hrule height 1pt}	
	\end{tabular*} 
	\end{specialtable}
\vspace{-6pt}

\begin{specialtable}[H] 
\widetable
\caption{\label{results_table_6}Numerical values of the $z$-component of the rescaled force $\tilde{\Phi}_{z}\equiv \partial \tilde\Phi/\partial\tilde z$ as well as the numbers $n_{\exp}, n_{\cos}$ and $n_{\mathrm{mix}}$ for points $A_1, A_2, B_1, B_2, C_1$ and $C_2$ for $\tilde\lambda_{\mathrm{eff}}=1$ and $\tilde\lambda_{\mathrm{eff}}=3$ in the left and right tables, respectively.}
\begin{tabular*}{\hsize}{@{}@{\extracolsep{\fill}}cccccccc|cccccccc@{}}
		\noalign{\hrule height 1pt}
		& $\bm{\tilde x}$ & $\bm{\tilde y}$ & $\bm{\tilde z}$ &$\bm{\tilde{\Phi}_z}$ & $\bm{n_{\exp}}$ & $\bm{n_{\cos}}$ & $\bm{n_{\mathrm{mix}}}$&& $\bm{\tilde x}$ & $\bm{\tilde y}$ & $\bm{\tilde z}$ &$\bm{\tilde{\Phi}_z}$ & $\bm{n_{\exp}}$ & $\bm{n_{\cos}}$ & $\bm{n_{\mathrm{mix}}}$\\	
		\noalign{\hrule height 0.5pt}
		$A_1$ & 0.5 & 0 & 0.5 & $-3.571$ & 85 & 21 & 15&$A_1$ & 0.5 & 0 & 0.5 &  $-5.072$ & 444 & 21 & 11\\
		$A_2$ & 0.5 & 0 & 0.1 &  $-1.673$ & 64 & 900 & 15&$A_2$ & 0.5 & 0 & 0.1 &  $-2.037$ & 331 & 863 & 13\\
		$B_1$ & 0.1 & 0 & 0.5 &  $-5.045$ & 74 & 20 & 13&		$B_1$ & 0.1 & 0 & 0.5 &  $-6.593$ & 397 & 19 & 9\\
		$B_2$ & 0.1 & 0 & 0.1 &  $-35.48$ & 15 & 444 & 5&$B_2$ & 0.1 & 0 & 0.1 &  $-36.04$ & 57 & 444 & 8\\
		$C_1$ & 0 & 0 & 0.5 &  $-5.241$ & 73 & 26 &  13&$C_1$ & 0 & 0 & 0.5 &  $-6.795$ & 392 & 24 & 12\\
		$C_2$ & 0 & 0 & 0.1 &  $-99.97$ & 8 & 678 & 4&$C_2$ & 0 & 0 & 0.1 &  $-100.7$ & 21 & 677 & 7\\
		\noalign{\hrule height 1pt}
		
	\end{tabular*}
	\end{specialtable}
\vspace{-6pt}

\vspace{-6pt}

\vspace{-6pt} 
\begin{paracol}{2}

\switchcolumn

\vspace{-6pt}
\section{Conclusions}\label{sec:5}
In this work, we have studied how the chimney topology $T\times T\times  R$ of the Universe affects the form of the gravitational potential and, consequently, that of the gravitational force. In this connection, we have proposed three alternative forms for each of the solutions. One of them (see  Equation~\rf{2.17}) relies on the Fourier expansion of the delta functions into series while using periodicity in two toroidal dimensions in the model. The second one (see Equation~\rf{2.18}) follows from the summation of solutions of the Helmholtz equation, each in the form of the Yukawa potential, for a source mass and all of its periodic images. Finally, the third form of the potential (see  Equation~\rf{2.20}) is formulated via the Ewald sums for Yukawa potentials. Subsequently, we have presented three alternative forms of the gravitational force (see  Equations~\rf{4.1}--\rf{4.3} and \rf{4.5}--\rf{4.7} for the $x$- and $z$-components, respectively) derived from the potential expressions.

In all three alternative forms, the screening length $\tilde\lambda_{\mathrm{eff}}$ serves as a crucial parameter, as it specifies the distance (from the source or the periodic images) where the gravitational potential undergoes exponential cutoff. This fact is most clearly demonstrated in the case where the solution takes on the form of summed Yukawa potentials \rf{2.18}.
The observational data show that this parameter should be less than 1 ($\tilde\lambda_{\mathrm{eff}} < 1$) in today's Universe. 

One of the main goals of this work was to reveal which of the obtained alternative formulas would serve better as a tool to be employed in numerical calculations. Namely, to show which formula would require less terms in the series to reach the desired precision. Our calculations have demonstrated that, for both the gravitational potentials and forces, two expressions involving plain summations of Yukawa potentials 
are preferable for the physically significant case $\tilde\lambda_{\mathrm{eff}} < 1$. 
However, in the case $\tilde\lambda_{\mathrm{eff}} \gtrsim 1$, the Yukawa--Ewald presentation stands out as the best alternative. 

Additionally, we have produced Figures~\ref{fig:1}--\ref{fig:10} for $\tilde\lambda_{\mathrm{eff}}= 0.01, 0.1, 1, 3$ to provide graphical demonstration of the gravitational potentials and force projections.

\authorcontributions{Conceptualization, M.E.; methodology, M.E., E.C. and A.Z.; formal analysis, M.E., A.M.II, E.C., M.B. and A.Z.; investigation, M.E., A.M.II, E.C., M.B. and A.Z.; writing---original draft preparation, A.Z.; writing---review and editing, M.E. and E.C.; visualization, M.E. and A.M.II; supervision, M.E. and A.Z.; project administration, M.E.; funding acquisition, M.E. All authors have read and agreed to the published version of the manuscript.} 

\funding{The work of M.E. and A.M.II was supported by the National Science Foundation HRD Award number 1954454.}

\conflictsofinterest{The authors have no conflicts of interest to declare that are relevant to the content of this article.} 

\end{paracol}
\reftitle{References}

%


\end{document}